\documentclass[a4paper, onecolumn]{quantumarticle}
\pdfoutput=1

\usepackage{contextuality_env}
\usepackage{orcidlink}
\usepackage{minted}
\usepackage[a4paper, margin=1in]{geometry}
\usepackage{tabularx}
\usepackage[
	style=numeric-comp,
	sorting=none,
	backend=biber
]{biblatex}
\usepackage[labelformat=simple]{subcaption}
\usepackage{cprotect}
\usepackage{hyperref}
\usepackage[section]{placeins}

\makeatletter
\AtBeginDocument{%
  \expandafter\renewcommand\expandafter\subsection\expandafter{%
    \expandafter\@fb@secFB\subsection
  }%
}
\makeatother

\definecolor{verylightgrey}{HTML}{F1F1F1}

\newcommand{\contextuality}{{\fontfamily{cmtt}\selectfont ConteXtu$\mathrm{\Lambda}$lity}}

\newminted{python}{
      frame=lines,
      framesep=2mm,
      baselinestretch=1.2,
      linenos=true,
      bgcolor=verylightgrey,
      fontsize=\footnotesize,
      tabsize=4,
      breaklines,
      autogobble,
}

\begin{document}
	\title{ConteXtuAlity: an open source Python package for contextuality}
	
	\author{Kim Vallée}
	\affiliation{Sorbonne Université, CNRS, LIP6, Paris 75005, France}
	\email{kim.vallee@protonmail.com}
	\orcid{0000-0002-4743-3984}
	
	\date{September 15, 2026}
	
	\maketitle
	
	\begin{abstract}
		We present an open source Python package, \contextuality{}, for computing various quantities related to contextuality. 
		This package features an interface to the sheaf theoretic framework for contextuality, providing an implementation of measurement scenarios and empirical models as well as linear programs to compute the contextual fraction and the signalling fraction.
		In this work, we demonstrate the package's features and give examples of its use. We also provide benchmarks, demonstrating how performances scale with the size of measurement scenarios. 
		The package is designed for both experienced theorists and newcomers to the field of contextuality alike, with an emphasis on flexibility and extensibility. 
	\end{abstract}
	
	\clearpage

	\tableofcontents
	
	\clearpage

\section{Introduction} \label{sec:introduction}

\subsection{Motivation} \label{ssec:motivation}

Physicists increasingly rely on computer programs as tools to simulate or predict behaviours.  Therefore, it is crucial to keep developing open-source software in order to make methods and results more consistent, reproducible and accessible.
Regarding quantum theory, some well-known pieces of software include the Python package QuTiP~\cite{lambert2025QuTiP5Quantum}, and its Julia equivalent QuantumToolbox.jl~\cite{mercurio2025QuantumToolboxjlEfficientJulia}, Ket.jl~\cite{araujo2025KetjlJuliaToolbox}, ProjectQ~\cite{steiger2018ProjectQOpenSource} or $\ket{\text{toqito}}$\cite{russo2020ToqitoTheoryQuantum} to name a few.
One particularly appreciated tool for quantum information is mathematical optimization, which has already proven useful in many areas, for instance through Semi-Definite Programming (SDP)~\cite{skrzypczyk2023SemidefiniteProgrammingQuantum}, but also in contextuality~\cite{schmid2018AllNoncontextualityInequalities,abramsky2017ContextualFractionMeasure,selby2022OpensourceLinearProgram,vallee2024CorrectedBellNoncontextuality}.

However, there is currently no unified package focused on contextuality, and extending the aforementioned packages would not make sense, as they focus on quantum theory, and research in contextuality often aims to be theory agnostic. Developing such a package would help the beginner contextuality enthusiast to build intuition by computing simple properties and exploring the concepts, but it would also help the advanced theorist who wishes to test new hypotheses and build a novel theory. It would remove much of the difficulty of implementing mathematical optimization problems that are already well known, in order to focus on the research that matters.

For this reason, we present \contextuality{}, an open-source Python package providing a practical interface for calculations and manipulations involving objects related to contextuality. Python was chosen for both its large community and its accessible approach to programming. 
This package relies on the sheaf theoretic approach to contextuality~\cite{abramsky2011SheaftheoreticStructureNonlocality}, an introduction of which can be found in Appendix~\ref{sec:sheaf_theoretic_framework_for_contextuality}, and we recommend the readers unfamiliar with this framework to read it.
In the future, we would like to extend this package to include many other frameworks, such as generalized contextuality~\cite{spekkens2005ContextualityPreparationsTransformations}, graph-theoretic approaches~\cite{cabello2014GraphTheoreticApproachQuantum,acin2015CombinatorialApproachNonlocality} and others.

\subsection{Package presentation} \label{ssec:package_presentation}

\begin{figure}[htpb]
	\centering
	\begin{subfigure}[b]{0.8\textwidth}
		\centering
		\includegraphics[width=0.8\textwidth]{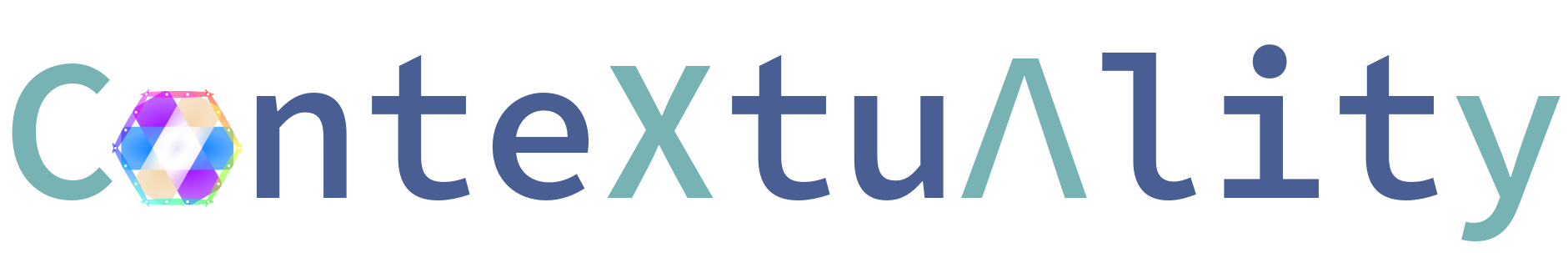}
		\caption{Full logo}
		\label{fig:full_logo}
	\end{subfigure}%
	\begin{subfigure}[b]{0.2\textwidth}
		\centering
		\includegraphics[width=0.5\textwidth]{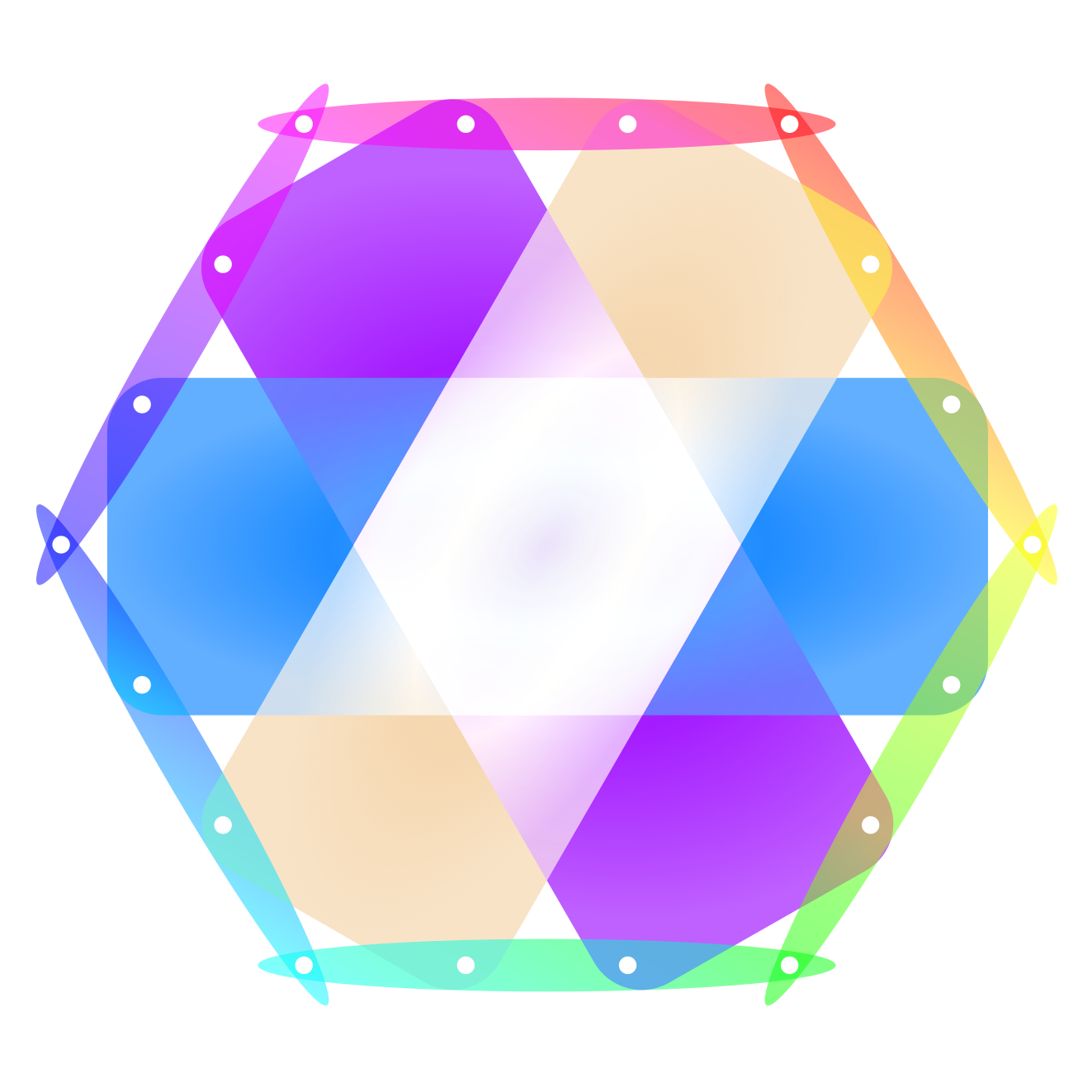}
		\caption{Icon logo}
		\label{fig:icon_logo}
	\end{subfigure}
	\caption{Two versions of the logo for the package.}
	\label{fig:logos}
\end{figure}

The Python package \contextuality{} is available on GitHub~\cite{vallee2025KimValleeContextualityGithub} and PyPi~\cite{vallee2025ContextualityPackagePyPi} and requires Python 3.11 or above. It is licensed with GNU GPLv3~\cite{gplv3}.
In addition, as the package uses mathematical optimization programs, one must install a solver; popular options include MOSEK~\cite{mosek} or HiGHS~\cite{schwendinger2025HighsHiGHSOptimization}, as well as the Python package \texttt{pycddlib}~\cite{GettingStartedPycddlib}. Full installation instructions can be found in the documentation available on ReadTheDocs~\cite{vallee2025RTDContextuality} and directly from the GitHub repository~\cite{vallee2025KimValleeContextualityGithub}. The package comes with two logos, a full logo and an icon logo represented in Figure~\ref{fig:logos}. At the time of writing the package is in version 2.0.2, and this is the version used in all the paper listings. 

In its current version, the package is made up of two main modules, the \verb|empirical_model| module and the \verb|measurement_scenario| module. In addition, it comprises other miscellaneous modules which are experimental. Only the \verb|utils| module is discussed here. These modules are presented in Section~\ref{sec:core_features}.

	\section{Core features} \label{sec:core_features}

The package is centred around two classes: \verb|MeasurementScenario| and \verb|EmpiricalModel|, which represent measurement scenarios and empirical models of the sheaf theoretic approach to contextuality, respectively. In this section we delve into these two classes, exposing their methods and variables that are useful for most users. A more detailed documentation can be found online~\cite{vallee2025RTDContextuality}. 

\subsection{The \texttt{MeasurementScenario} class} \label{ssec:the-texttt-measurementscenario_class}

In the sheaf-theoretic approach to contextuality, any abstract setup can be represented by a measurement scenario $\XMO$\footnote{We refer the reader to Section~\ref{ssec:overview_of_sheaf_theoretic_contextuality}.}.
Such scenarios are represented by instances of the \texttt{MeasurementScenario} class, which require three arguments:
\begin{itemize}
	\item \texttt{X} for the set of measurements;
	\item \texttt{M} for the set of contexts;
	\item and \texttt{O} for the set of outcomes.
\end{itemize}
These arguments may be integers or strings, but strings will be converted to symbols using the SymPy module~\cite{meurer2017SymPySymbolicComputing}. We start with a brief example of a valid definition of \texttt{MeasurementScenario} in Listing~\ref{lst:measurement_scenario}.

\begin{listing}[!ht]
\begin{pythoncode}
from contextuality import MeasurementScenario

X = ["A0", "A1", "B0", "B1"]
M = [["A0", "B0"], ["A0", "B1"], ["A1", "B0"], ["A1", "B1"]]
O = [0, 1]
scenario = MeasurementScenario(X, M, O)
print(scenario)
# OUT: MeasurementScenario(X=[A0, A1, B0, B1], M=[[A0, B0], [A0, B1], [A1, B0], [A1, B1]], O=[0, 1]) 
\end{pythoncode}
\caption{Definition of a measurement scenario.}
\label{lst:measurement_scenario}
\end{listing}

Building a measurement scenario is the first step, and in itself the \texttt{MeasurementScenario} class already provides a wide range of useful methods and variables summarized in Table~\ref{tab:list-of-methods-MeasurementScenario}.

\bgroup
\renewcommand*{\arraystretch}{1.5}
\begin{table}[htpb]
	\centering
	\begin{tabularx}{\textwidth}{@{} p{\dimexpr.3\linewidth-\tabcolsep} p{\dimexpr.7\linewidth-\tabcolsep}@{}}
		\toprule
		Method name & Description \\
		\midrule
		\texttt{X}, \texttt{M}, \texttt{O} & Accessors for the \texttt{X}, \texttt{M} and \texttt{O} variables given at initialization. They might not be of the same type, as strings are converted to symbols. \\
		\verb|outcomes_global()| & Method that returns all possible global assignments, in the order given by \texttt{X}. \\
		\verb|incidence_matrix| & Accessor for the incidence matrix associated to this measurement scenario. \\
		\verb|all_outcomes| & Accessor that gives all possible outcomes for a context. \\
		\verb|generate_deterministic(p)| & Method that generates a deterministic empirical model of the measurement scenario, with the unit probabilities positions given by the argument \verb|p|.\\
		\bottomrule
	\end{tabularx}
	\caption{Non-exhaustive list of methods and accessors in the \texttt{MeasurementScenario} class.}
	\label{tab:list-of-methods-MeasurementScenario}
\end{table}
\egroup

A major advantage of the \verb|incidence_matrix| accessor is that it automatically computes the matrix, a task that previously had to be performed case by case. In turn the incidence matrix is very useful for computations of the contextual fraction and signalling fraction which we will see in Section~\ref{ssec:the-texttt-empiricalmodel_class}.

\subsection{The \texttt{EmpiricalModel} class} \label{ssec:the-texttt-empiricalmodel_class}

An empirical model $e$  is a family of probability distributions on the contexts of a measurement scenario $\XMO$. Within this package, it is represented by an instance of the \texttt{EmpiricalModel} class. 

There are few options for defining an empirical model in \contextuality{}. The family of probability distributions can be provided at its instantiation, assigned later, or it can be generated with a set of POVMs and a state. We summarize these cases with examples in Listing~\ref{lst:empirical_model_ex}.

\begin{listing}[!ht]
\begin{pythoncode}
from contextuality import MeasurementScenarioImplementations, EmpiricalModel

kcbs_scenario = MeasurementScenarioImplementations.kcbs()
em = EmpiricalModel(kcbs_scenario)
print(em)

# OUT: EmpiricalModel(MeasurementScenario(X=[0, 1, 2, 3, 4], M=[[0, 1], [1, 2], [2, 3], [3, 4], [4, 0]], O=[0, 1]))

################# OR #################

simple_data = [[1,0,0,0] * 5]
em = EmpiricalModel(kcbs_scenario, simple_data)
print(em)

# OUT: EmpiricalModel(MeasurementScenario(X=[0, 1, 2, 3, 4], M=[[0, 1], [1, 2], [2, 3], [3, 4], [4, 0]], O=[0, 1])
# 	1.00 0.00 0.00 0.00 
# 	...
# )

################# OR #################

em = EmpiricalModel(kcbs_scenario)
em.vector = simple_data
print(em)

# OUT: EmpiricalModel(MeasurementScenario(X=[0, 1, 2, 3, 4], M=[[0, 1], [1, 2], [2, 3], [3, 4], [4, 0]], O=[0, 1])
# 	1.00 0.00 0.00 0.00 
#       ...
# )

################# OR #################

# see notebooks/KCBS.ipynb for the full quantum realization example
rho = ...
PVMs = ...
em = EmpiricalModel(kcbs_scenario)
em.quantum_realisation(rho, PVMs)
\end{pythoncode}
\caption{The various ways to define an empirical model.}
\label{lst:empirical_model_ex}
\end{listing}

The \verb|EmpiricalModel| class comes equipped with a wide range of methods and attributes. We expose some of these in Table~\ref{tab:list-of-methods-EmpiricalModel} and the full set of available methods can be found in the documentation~\cite{vallee2025RTDContextuality}.
\bgroup
\renewcommand*{\arraystretch}{1.5}
\begin{table}[htpb]
	\centering
	\begin{tabularx}{\textwidth}{@{} p{\dimexpr.4\linewidth-\tabcolsep} p{\dimexpr.6\linewidth-\tabcolsep}@{}}
		\toprule
		Method name & Description \\
		\midrule
		\verb|vector| & Accessor and setter for the vector array.\\
		\verb|mvector| & Accessor for the matrix version of the vector, where each row represents one probability distribution in one context. \\
		\verb|is_valid| & Property that is true iff the empirical model is a valid family of probability distributions. \\
		\verb|quantum_realisation(rho, pvms)| & Generates the vector for the current empirical model from a state \verb|rho| and a set of projective measurements  \verb|pvms|. \\
		\verb|compute_sf(solver="MOSEK",| \par\hfill\verb|verbose=False)| & Computes the signalling fraction of the empirical model.\\
		\verb|compute_cf(solver="MOSEK",|\par\hfill\verb|verbose=False)| & Computes the contextual fraction of the empirical model.\\
		\bottomrule
	\end{tabularx}
	\caption{Non-exhaustive list of methods and accessors in the \texttt{EmpiricalModel} class.}
	\label{tab:list-of-methods-EmpiricalModel}
\end{table}
\egroup
In addition, the \verb|EmpiricalModel| class contains special methods for multiplying and adding empirical models together, as well as multiplying and dividing them with floating-point numbers. We showcase their usage in Listing~\ref{lst:empirical_model_special_methods}.

\begin{listing}[!ht]
\begin{pythoncode}
from contextuality import MeasurementScenarioImplementations, EmpiricalModel

kcbs_scenario = MeasurementScenarioImplementations.kcbs()
em1 = EmpiricalModel(kcbs_scenario, [[1,0,0,0] * 5])
em2 = EmpiricalModel(kcbs_scenario, [[0,0,0,1] * 5])

print(0.5 * em1)
# OUT: EmpiricalModel(MeasurementScenario(X=[0, 1, 2, 3, 4], M=[[0, 1], [1, 2], [2, 3], [3, 4], [4, 0]], O=[0, 1])
#	0.50 0.00 0.00 0.00 
# 	0.50 0.00 0.00 0.00 
# 	0.50 0.00 0.00 0.00 
# 	0.50 0.00 0.00 0.00 
# 	0.50 0.00 0.00 0.00 
# )

print(0.3 * em1 + 0.7 * em2)
# EmpiricalModel(MeasurementScenario(X=[0, 1, 2, 3, 4], M=[[0, 1], [1, 2], [2, 3], [3, 4], [4, 0]], O=[0, 1])
# 	0.30 0.00 0.00 0.70 
# 	0.30 0.00 0.00 0.70 
# 	0.30 0.00 0.00 0.70 
# 	0.30 0.00 0.00 0.70 
# 	0.30 0.00 0.00 0.70 
# )
\end{pythoncode}
\caption{Examples of the special methods for the \texttt{EmpiricalModel} class.}
\label{lst:empirical_model_special_methods}
\end{listing}

\subsection{Miscellaneous} \label{ssec:miscellaneous}

\subsubsection{The \texttt{MeasurementScenarioImplementations} class} \label{sssec:the-texttt-measurementscenarioimplemgentation_class}

There are some well-known measurement scenarios, which are often used in the literature, such as the Bell or Clauser-Horne-Shimony-Holt (CHSH) scenario~\cite{bell1964EinsteinPodolskyRosen,clauser1969ProposedExperimentTest}, the Klyachko-Can-Binicioğlu-Shumovsky (KCBS) scenario~\cite{klyachko2008SimpleTestHidden} and the Peres-Mermin square~\cite{peres1991TwoSimpleProofs,mermin1990SimpleUnifiedForm}.
For convenient access, these scenarios are pre-defined within the \texttt{MeasurementScenarioImplementations} class through static methods. An example usage is given in Listing~\ref{lst:msi_class_examples}.

\begin{listing}[!ht]
\begin{pythoncode}
from contextuality import MeasurementScenarioImplementations

kcbs_scenario = MeasurementScenarioImplementations.kcbs()
print(kcbs_scenario)
# OUT: MeasurementScenario(X=[0, 1, 2, 3, 4], M=[[0, 1], [1, 2], [2, 3], [3, 4], [4, 0]], O=[0, 1])
chsh_scenario = MeasurementScenarioImplementations.chsh()
print(chsh_scenario)
# OUT: MeasurementScenario(X=[0, 1, 2, 3], M=[[0, 2], [0, 3], [1, 2], [1, 3]], O=[0, 1])
pm_scenario = MeasurementScenarioImplementations.peres_mermin()
print(pm_scenario)
# OUT: MeasurementScenario(X=[0, 1, 2, 3, 4, 5, 6, 7, 8], M=[[0, 1, 2], [3, 4, 5], [6, 7, 8], [0, 3, 6], [1, 4, 7], [2, 5, 8]], O=[0, 1])
\end{pythoncode}
\caption{Examples of usage of the \texttt{MeasurementScenarioImplementations} class.}
\label{lst:msi_class_examples}
\end{listing}

\subsubsection{The \texttt{utils} module} \label{sssec:the_utils_module}

There is an additional module called \texttt{utils} that contains mostly experimental and legacy functions, and in particular it contains functions for constructing the polytopes associated with measurement scenarios, both the non-contextual polytope and the signalling polytope. Some of these functions are described in Table~\ref{tab:list-of-methods-utils}.

\bgroup
\renewcommand*{\arraystretch}{1.5}
\begin{table}[htpb]
	\centering
	\begin{tabularx}{\textwidth}{@{} p{\dimexpr.45\linewidth-\tabcolsep} p{\dimexpr.55\linewidth-\tabcolsep}@{}}
		\toprule
		Function name & Description \\
		\midrule
		\verb|nc_polytope(MS, representation="V")| & Generates the polytope for the measurement scenario  \texttt{MS} in the \texttt{representation}, which can be  \texttt{V}(ertex) or \texttt{H}(alfspace). The \texttt{H} representation is not yet optimized.\\
		\verb|signalling_polytope(MS,|\par\hspace{4.5em}\verb|include_NS_polytope=True)| & Similar to \verb|nc_polytope| but generates the signalling polytope in the \texttt{V} representation only. If \verb|include_NS_polyope| is set to \texttt{False} then the non-contextual extremal points are excluded.\\
		\verb|compute_max_cf(MS, sigma, eta,|\par\hspace{4.5em}\verb|big_m=2, solver="MOSEK",|\par\hspace{4.5em}\verb|verbose=False)| & Experimental function to compute the maximum CF attainable with parameters \texttt{sigma} and  \texttt{eta}. It uses linear programming and the \verb|big_m| method. It was used before the analytical result of~\cite{vallee2024CorrectedBellNoncontextuality} was derived.\\
		\bottomrule
	\end{tabularx}
	\caption{Non-exhaustive list of functions in the \texttt{utils} module.}
	\label{tab:list-of-methods-utils}
\end{table}
\egroup

	\section{Example usage} \label{sec:example_usage}

We now turn to examples illustrating how \contextuality{} can be used in a variety of situations. 
In this section, we examine two experimental studies, one by Lapkiewicz et al.~\cite{lapkiewicz2011ExperimentalNonclassicalityIndivisible} and one by Wang et al.~\cite{wang2022SignificantLoopholefreeTest}. We then show how the package can be used to sample the contextual and signalling fractions in the Bell-CHSH scenario, before demonstrating how more restrictive hidden variable models can also be considered.

\subsection{Lapkiewicz et al. experimental setup} \label{ssec:reference-cite-lapkiewicz2011experimentalnonclassicalityindivisible-}

We start with the work of Lapkiewicz et al.~\cite{lapkiewicz2011ExperimentalNonclassicalityIndivisible} that investigates the KCBS scenario~\cite{klyachko2008SimpleTestHidden}. This scenario $\XMO[\text{KCBS}]$ is composed of:
\begin{subequations}
	\begin{align}
		X_{\text{KCBS}} &= \left\{A_1, A_2, A_3, A_4, A_5\right\} \\
		\mathcal{M}_{\text{KCBS}} &= \left\{A_1A_2, A_2A_3,A_3A_4,A_4A_5,A_5A_1\right\} \\
		O_{\text{KCBS}} &= \left\{-1,1\right\} 
	\end{align}
\end{subequations}
This scenario is best known for the following inequality defining a facet of the non-contextual polytope: 
\begin{equation} \label{eq:KCBS_ineq}
\braket{A_1A_2} + \braket{A_2 A_3} + \braket{A_3 A_4} + \braket{A_4 A_5} + \braket{A_5 A_1} \ge -3
.\end{equation}
This inequality is violated by quantum theory for a known set of observables and a state given in Refs.~\cite{cabello2010NonContextualityPhysicalTheories,araujo2013AllNoncontextualityInequalities}. The empirical model $e_{\text{KCBS}}$ obtained from the experimental data is given in Table~\ref{tab:empirical_model_lapkiewicz}. We first show how to compute the signalling fraction and the contextual fraction from this empirical model, treating it as a standard KCBS scenario, in Listing~\ref{lst:cf_sf_lapkiewicz}.

\begin{table}[htpb]
	\centering
	\begin{tabular}{@{}l c c c c@{}} \toprule
		& $-1,-1$ & $-1,+1$ &  $+1,-1$ & $+1,+1$ \\ \midrule
		$A_1A_2$ & $0$ & $0.471$ & $0.432$ & $0.097$ \\
		$A_2A_3$ & $0$ & $0.429$ & $0.473$ & $0.098$ \\
		$A_3A_4$ & $0$ & $0.429$ & $0.426$ & $0.146$ \\
		$A_4A_5$ & $0$ & $0.466$ & $0.439$ & $0.095$\\
		$A_5A'_1$ & $0$ & $0.414$ & $0.469$ & $0.117$\\
		\bottomrule
	\end{tabular}
	\caption{Experimental data from Ref~\cite{lapkiewicz2011ExperimentalNonclassicalityIndivisible}, gathered in an empirical model $e_{\text{KCBS}}$.}
	\label{tab:empirical_model_lapkiewicz}
\end{table}

\begin{listing}[!ht]
\begin{pythoncode}
from contextuality import MeasurementScenarioImplementations, EmpiricalModel

ms = MeasurementScenarioImplementations.kcbs()

em_vector = [
    0, 0.471, 0.432, 0.097,
    0, 0.429, 0.473, 0.098,
    0, 0.429, 0.426, 0.146,
    0, 0.466, 0.439, 0.095,
    0, 0.414, 0.469, 0.117,
]

em = EmpiricalModel(ms, em_vector)

sf = em.compute_sf()['SF']
cf = em.compute_cf()['CF']

print(f"CF: {round(cf,4)}")
# OUT: 'CF: 0.447'
print(f"SF: {round(sf,4)}")
# OUT: 'SF: 0.065'
\end{pythoncode}
\caption{Listing to compute the contextual fraction and the signalling fraction from the data of~\cite{lapkiewicz2011ExperimentalNonclassicalityIndivisible} assuming a standard KCBS scenario.}
\label{lst:cf_sf_lapkiewicz}
\end{listing}

We can relate the contextual fraction obtained in Listing~\ref{lst:cf_sf_lapkiewicz} to the violation of Equation~\eqref{eq:KCBS_ineq} using the following equation from Ref.~\cite{abramsky2017ContextualFractionMeasure}\footnote{This assumes that inequality~\eqref{eq:KCBS_ineq} is the one maximally violated with the empirical model of Table~\ref{tab:empirical_model_lapkiewicz}}:
\begin{equation} \label{eq:cf_from_ineq}
\CF(e) = \max \left( 0, \frac{\beta - \beta_{\text{cl}}}{\beta_{\text{max}} - \beta_{\text{cl}}}\right)  
.\end{equation} 
Where $\beta_{\text{cl}}$ is the classical bound of the inequality, $\beta_{\text{max}}$ is the algebraic bound, and $\beta$ is the value obtained experimentally. Rearranging the terms, as we have $\beta_{\text{max}} - \beta_{\text{cl}} \neq 0$, we obtain:
\begin{equation} 
\beta = \CF(e) \left[ \beta_{\text{max}} - \beta_{\text{cl}} \right] + \beta_{\text{cl}} 
.\end{equation} 
Therefore, we conclude that $\beta \approx -3.894$, corresponding to the result given in Ref~\cite{lapkiewicz2011ExperimentalNonclassicalityIndivisible}, that is $\beta \approx -3.893$. The slight difference might be due to rounding error in the data.

In the experimental setup of Ref.~\cite{lapkiewicz2011ExperimentalNonclassicalityIndivisible}, the observable $A_1$ in context $A_1 A_2$ is implemented differently from the one in the context $A_5 A_1$. To account for this difference we can consider a modified version, with the following alternative measurement scenario $\XMO[\text{KCBS}']$:
\begin{subequations}
	\begin{align}
		X_{\text{KCBS}'} &= \left\{A_1, A_2, A_3, A_4, A_5, A_1'\right\} \\
		\mathcal{M}_{\text{KCBS}'} &= \left\{A_1A_2, A_2A_3, A_3A_4, A_4A_5, A_5A_1'\right\} \\
		O_{\text{KCBS}'} &= \left\{-1, 1\right\}
	\end{align}
\end{subequations}
Where we added a new observable $A_1'$ unrelated to $A_1$, thus changing the set of contexts $\mathcal{M}_{\text{KCBS}'}$. 
As expected, treating $A_1$ and $A_1'$ as distinct observables changes both the contextual and signalling fraction, as given by the Listing~\ref{lst:cf_sf_lapkiewicz_2}, illustrating how the package can accommodate for different measurement scenarios.

\begin{listing}[!ht]
\begin{pythoncode}
from contextuality import MeasurementScenario, EmpiricalModel

ms = MeasurementScenario(X=["A1", "A2", "A3", "A4", "A5", "A1prime"],
                         M=[["A1", "A2"], ["A2", "A3"], ["A3", "A4"], ["A4", "A5"], ["A5", "A1prime"]],
                         O=[0, 1])

em_vector = [
    0, 0.471, 0.432, 0.097,
    0, 0.429, 0.473, 0.098,
    0, 0.429, 0.426, 0.146,
    0, 0.466, 0.439, 0.095,
    0, 0.414, 0.469, 0.117,
]

em = EmpiricalModel(ms, em_vector)

sf = em.compute_sf()['SF']
cf = em.compute_cf()['CF']

print(f"CF: {round(cf,4)}")
# OUT: 'CF: 0.065'
print(f"SF: {round(sf,4)}")
# OUT: 'SF: 0.065'
\end{pythoncode}
\caption{Listing to compute the contextual fraction and the signalling fraction considering different observables in context $A_1A_2$ and $A_5A_1$ from the data of~\cite{lapkiewicz2011ExperimentalNonclassicalityIndivisible}.}
\label{lst:cf_sf_lapkiewicz_2}
\end{listing}

\subsection{Wang et al. experimental setup} \label{ssec:contextual_fraction_of_reference-cite-wang2022significantloopholefreetest-}

The work of Wang et al.~\cite{wang2022SignificantLoopholefreeTest} is a practical example for demonstrating another way to compute the contextual and signalling fractions of an empirical model from realistic data.
It implements a contextuality test that resembles the CHSH scenario~\cite{clauser1969ProposedExperimentTest} given in Equations~\eqref{eqs:CHSH_scenario}. The results of the experiment are the averages of the observables. We report the results of the experiment in Table~\ref{tab:empirical_model_chsh_wang}.

\begin{table}[htpb]
	\centering
	\begin{tabular}{@{}l c c c c@{}} \toprule
		& $-1,-1$ & $-1,+1$ &  $+1,-1$ & $+1,+1$ \\ \midrule
		$O_0O_1$ & $0.4313$ & $0.0683$ & $0.1235$ & $0.3769$ \\
		$O_1O_2$ & $0.4638$ & $0.0895$ & $0.0980$ & $0.3487$ \\
		$O_2O_3$ & $0.4778$ & $0.0900$ & $0.0761$ & $0.3561$ \\
		$O_3O_0$ & $0.1223$ & $0.4334$ & $0.3749$ & $0.0694$\\
		\bottomrule
	\end{tabular}
	\caption{Empirical model corresponding to the experimental data obtained in Reference~\cite[Table 1]{wang2022SignificantLoopholefreeTest}, where the correlations have been converted to probabilities using the equations provided in~\cite[Eq. 3]{araujo2013AllNoncontextualityInequalities}.}
	\label{tab:empirical_model_chsh_wang}
\end{table}

Similarly to Listing~\ref{lst:cf_sf_lapkiewicz}, in Listing~\ref{lst:cf_and_sf_wang2022} we compute the contextual and signalling fractions of the empirical model given in Table~\ref{tab:empirical_model_chsh_wang}. A few more lines of code are required to change the correlations into a valid empirical model, the associated method is explained in detail in~\cite{araujo2013AllNoncontextualityInequalities}. A further analysis, using Equation~\eqref{eq:cf_from_ineq}, reveals that this gives a violation of the inequality $\beta \approx 2.5258$, which corresponds to the value obtained in the original paper~\cite{wang2022SignificantLoopholefreeTest}:  $\beta = 2.526 \pm 0.016$.

\begin{listing}[!ht]
\begin{pythoncode}
from contextuality import MeasurementScenarioImplementations, EmpiricalModel

# Correlation obtained from the experimental data
correlations = [
    [0.6164,-0.0008,0.1096],
    [0.625, 0.1066, 0.1236],
    [0.6678, 0.1356, 0.1078],
    [-0.6166, 0.1114, -0.0056]
]

# Convert correlations to probability vector using Araujo et al. formula
probability_vector = []
for cor in correlations:
    p00 = (1 + cor[0] + cor[1] + cor[2]) / 4
    p01 = (1 - cor[0] + cor[1] - cor[2]) / 4
    p10 = (1 - cor[0] - cor[1] + cor[2]) / 4
    p11 = (1 + cor[0] - cor[1] - cor[2]) / 4
    probability_vector.extend([p00, p01, p10, p11])

ms = MeasurementScenarioImplementations.chsh()
em = EmpiricalModel(ms, probability_vector)

sf = em.compute_sf()['SF']
cf = em.compute_cf()['CF']

print(f"CF: {round(cf,4)}")
# OUT: 'CF: 0.2629'
print(f"SF: {round(sf,4)}")
# OUT: 'SF: 0.0646'
\end{pythoncode}
\caption{Listing to compute the empirical model from the correlations, and then the contextual and signalling fraction from Ref.~\cite{wang2022SignificantLoopholefreeTest}.}
\label{lst:cf_and_sf_wang2022}
\end{listing}

This example shows how \contextuality{} allows to compute the contextual and signalling fractions in different scenarios once again, showcasing that the code is very accessible with very few steps.

\subsection{Colormap of the contextual fraction and signalling fraction} \label{ssec:colormap_of_the_contextual_fraction_and_signalling_fraction}

Our last example will combine the use of \verb|compute_cf()|, \verb|compute_sf()| and the addition and multiplication magic methods of an \verb|EmpiricalModel| presented in Section~\ref{ssec:the-texttt-empiricalmodel_class}. Our objective is to obtain a heat map of the contextual fraction and the signalling fraction in the Bell-CHSH measurement scenario.

First, let us define the Bell-CHSH measurement scenario $\XMO[\text{CHSH}]$ as:
\begin{subequations}
	\begin{align}
		X_{\text{CHSH}} &= \left\{A, B, A', B'\right\} \\
		\mathcal{M}_{\text{CHSH}} &= \left\{AB, AB', A'B, A'B'\right\} \\
		O_{\text{CHSH}} &= \left\{0, 1\right\}
	\end{align}
\end{subequations}
Additionally, we establish some empirical models for the Bell-CHSH scenario in Table~\ref{tab:empirical_models_CHSH}, which will serve as reference points in the final plot.
\begin{table}[htpb]
    \centering
    \begin{subtable}[t]{0.3\linewidth}
        \centering
        \begin{tabular}{@{}l c c c c@{}} \toprule
            & $00$ & $01$ & $10$ & $11$ \\ \midrule
            $AB$  & 1 & 0 & 0 & 0 \\
            $AB'$ & 1 & 0 & 0 & 0 \\
            $A'B$ & 1 & 0 & 0 & 0 \\
            $A'B'$& 0 & 1 & 0 & 0 \\
            \bottomrule
        \end{tabular}
        \caption{$e^{\text{MS}_1}$}
    \end{subtable}
    \hfill
    \begin{subtable}[t]{0.3\linewidth}
        \centering
        \begin{tabular}{@{}l c c c c@{}} \toprule
            & $00$ & $01$ & $10$ & $11$ \\ \midrule
            $AB$  & 0 & 0 & 0 & 1 \\
            $AB'$ & 0 & 0 & 0 & 1 \\
            $A'B$ & 0 & 0 & 0 & 1 \\
            $A'B'$& 0 & 0 & 1 & 0 \\
            \bottomrule
        \end{tabular}
        \caption{$e^{\text{MS}_2}$}
    \end{subtable}
    \hfill
    \begin{subtable}[t]{0.3\linewidth}
        \centering
        \begin{tabular}{@{}l c c c c@{}} \toprule
            & $00$ & $01$ & $10$ & $11$ \\ \midrule
            $AB$  & $\sfrac{1}{2}$ & 0 & 0 & $\sfrac{1}{2}$ \\
            $AB'$ & $\sfrac{1}{2}$ & 0 & 0 & $\sfrac{1}{2}$ \\
            $A'B$ & $\sfrac{1}{2}$ & 0 & 0 & $\sfrac{1}{2}$ \\
            $A'B'$& 0 & $\sfrac{1}{2}$ & $\sfrac{1}{2}$ & 0 \\
            \bottomrule
        \end{tabular}
        \caption{$e^{\text{PR}_1}$}
    \end{subtable}
    \\[1ex]
    \begin{subtable}[t]{0.3\linewidth}
        \centering
        \begin{tabular}{@{}l c c c c@{}} \toprule
            & $00$ & $01$ & $10$ & $11$ \\ \midrule
            $AB$  & 0 & 0 & 1 & 0 \\
            $AB'$ & 0 & 0 & 1 & 0 \\
            $A'B$ & 0 & 0 & 1 & 0 \\
            $A'B'$& 0 & 0 & 0 & 1 \\
            \bottomrule
        \end{tabular}
        \caption{$e^{\text{MS}_3}$}
    \end{subtable}
    \hfill
    \begin{subtable}[t]{0.3\linewidth}
        \centering
        \begin{tabular}{@{}l c c c c@{}} \toprule
            & $00$ & $01$ & $10$ & $11$ \\ \midrule
            $AB$  & 0 & 1 & 0 & 0 \\
            $AB'$ & 0 & 1 & 0 & 0 \\
            $A'B$ & 0 & 1 & 0 & 0 \\
            $A'B'$& 1 & 0 & 0 & 0 \\
            \bottomrule
        \end{tabular}
        \caption{$e^{\text{MS}_4}$}
    \end{subtable}
    \hfill
    \begin{subtable}[t]{0.3\linewidth}
    \centering
    \begin{tabular}{@{}l c c c c@{}} \toprule
	& $00$ & $01$ & $10$ & $11$ \\ \midrule
	$AB$  & 0 & $\sfrac{1}{2}$ & $\sfrac{1}{2}$ & 0 \\
	$AB'$ & 0 & $\sfrac{1}{2}$ & $\sfrac{1}{2}$ & 0 \\
	$A'B$ & 0 & $\sfrac{1}{2}$ & $\sfrac{1}{2}$ & 0 \\
	$A'B'$& $\sfrac{1}{2}$ & 0 & 0 & $\sfrac{1}{2}$ \\
	\bottomrule
    \end{tabular}
    \caption{$e^{\text{PR}_2}$}
    \end{subtable}
    \\[1ex]
    \begin{subtable}[t]{0.3\linewidth}
        \centering
        \begin{tabular}{@{}l c c c c@{}} \toprule
            & $00$ & $01$ & $10$ & $11$ \\ \midrule
            $AB$  & $\sfrac{1}{4}$ & $\sfrac{1}{4}$ & $\sfrac{1}{4}$ & $\sfrac{1}{4}$ \\
            $AB'$ & $\sfrac{1}{4}$ & $\sfrac{1}{4}$ & $\sfrac{1}{4}$ & $\sfrac{1}{4}$ \\
            $A'B$ & $\sfrac{1}{4}$ & $\sfrac{1}{4}$ & $\sfrac{1}{4}$ & $\sfrac{1}{4}$ \\
            $A'B'$& $\sfrac{1}{4}$ & $\sfrac{1}{4}$ & $\sfrac{1}{4}$ & $\sfrac{1}{4}$ \\
            \bottomrule
        \end{tabular}
        \caption{$e^{\text{MM}}$}
    \end{subtable}
    \caption{Empirical models in the CHSH scenario, where $\text{MS}$ stands for \textit{Maximally Signalling}, $\text{PR}$ refers to the \textit{Popescu-Rohrlich box}~\cite{popescu1994QuantumNonlocalityAxiom} and MM to the maximally mixed empirical model.}
    \label{tab:empirical_models_CHSH}
\end{table}

Note that we have the following relations between the empirical models:
\begin{align}
	\label{eq:PR1} e^{\text{PR}_1} &= \frac{1}{2} e^{\text{MS}_1} + \frac{1}{2} e^{\text{MS}_2} \\
	\label{eq:PR2} e^{\text{PR}_2} &= \frac{1}{2} e^{\text{MS}_3} + \frac{1}{2} e^{\text{MS}_4} \\
	\label{eq:MM} e^{\text{MM}} &= \frac{1}{2} e^{\text{PR}_1} + \frac{1}{2} e^{\text{PR}_2}
\end{align}
We can use relations~\eqref{eq:PR1} and~\eqref{eq:PR2} to establish a sample of the contextual fraction and the signalling fraction following a method close to bilinear approximation, i.e., let us define:
\begin{equation}
	\mathcal{E} = \left\{ x y e^{\text{MS}_1} + x (1-y) e^{\text{MS}_2}) + (1-x) y e^{\text{MS}_3} + (1 - x)(1 - y) e^{\text{MS}_4}) \middle| x, y \in \mathbb{R}^{+}, x \le 1, y \le 1  \right\}
.\end{equation} 
In Listing~\ref{lst:heatmap_cf_sf} we present how to compute the signalling fraction and contextual fraction with a discretized version of $\mathcal{E}$, namely $\mathcal{E}_{\ell}$:
\begin{equation} \label{eq:discretized_E}
	\mathcal{E}_{\ell} = \left\{ x y e^{\text{MS}_1} + x (1-y) e^{\text{MS}_2} + (1-x) y e^{\text{MS}_3} + (1 - x)(1 - y) e^{\text{MS}_4} \middle| x,y \in \left\{0, \frac{1}{\ell - 1}, \ldots, \frac{\ell - 2}{\ell - 1}, 1 \right\}  \right\}
.\end{equation} 
Here, we have chosen $\ell = 100$.

\begin{listing}[!ht]
\begin{pythoncode}
from contextuality import EmpiricalModel, MeasurementScenarioImplementations
import numpy as np

chsh_ms = MeasurementScenarioImplementations.chsh()

# List of Maximally Signalling empirical models
empirical_models = [
    EmpiricalModel(chsh, chsh.generate_deterministic([0,0,0,1])),
    EmpiricalModel(chsh, chsh.generate_deterministic([3,3,3,2])),
    EmpiricalModel(chsh, chsh.generate_deterministic([2,2,2,3])),
    EmpiricalModel(chsh, chsh.generate_deterministic([1,1,1,0]))
]

nb_points = 100

x_range = np.linspace(0, 1, nb_points)
y_range = np.linspace(0, 1, nb_points)

CFs = np.zeros((nb_points, nb_points))
SFs = np.zeros((nb_points, nb_points))

for i, y in enumerate(y_range):
    em1 = (1 - y) * empirical_models[0] + y * empirical_models[1]
    em2 = (1 - y) * empirical_models[2] + y * empirical_models[3]
    for j, x in enumerate(x_range):
        em = (1 - x) * p1 + x * p2
        CFs[j, i] = em.compute_cf()["CF"]
        SFs[j, i] = em.compute_sf()["SF"]
\end{pythoncode}
\caption{Computing the contextual fraction and signalling fraction of each point in the set $\mathcal{E}_{\ell = 100}$.}
\label{lst:heatmap_cf_sf}
\end{listing}

\begin{figure}[htpb]
	\centering
	\begin{subfigure}[b]{0.5\textwidth}
		\centering
		\includegraphics[width=\textwidth]{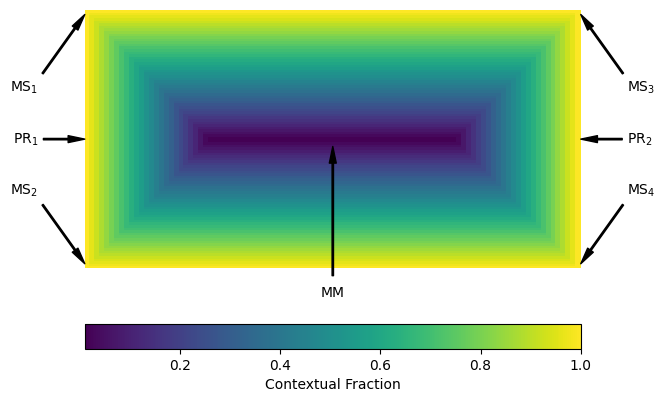}
		\caption{}
		\label{fig:figures-heatmap_cf}
	\end{subfigure}%
	\begin{subfigure}[b]{0.5\textwidth}
		\centering
		\includegraphics[width=\linewidth]{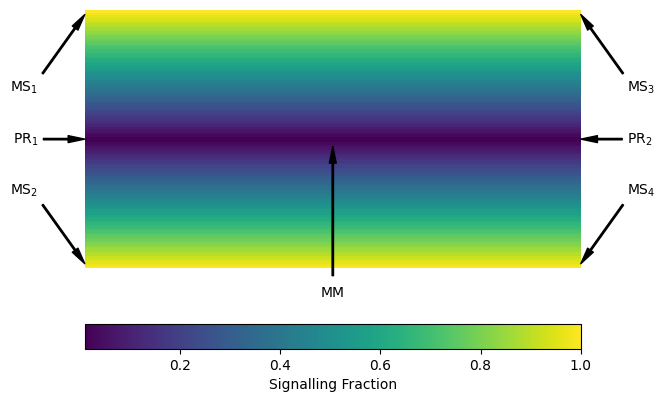}
		\caption{}
		\label{fig:figures-heatmap_sf}
	\end{subfigure}
	\caption{Heatmaps of \subref{fig:figures-heatmap_cf} the contextual fraction and \subref{fig:figures-heatmap_sf} the signalling fraction following the results of Listing~\ref{lst:heatmap_cf_sf}.}
	\label{fig:figures-heatmap}
\end{figure}

The results obtained by Listing~\ref{lst:heatmap_cf_sf} can be graphically represented on a heatmap, which we show in Figure~\ref{fig:figures-heatmap}.

\subsection{Restricting non-contextual behaviours} \label{ssec:restricting_non-contextual_behaviours}

Different contextuality frameworks have different constraints imposed on their hidden variable models. In the sheaf-theoretic framework for contextuality, the allowed hidden variable models are directly related to the incidence matrix~\cite{abramsky2011SheaftheoreticStructureNonlocality}. As a matter of fact, the incidence matrix columns can be understood as the extremal points of the non-contextual polytope~\cite{abramsky2017ContextualFractionMeasure}, thus, modifying the shape of the incidence matrix has direct consequences on the hidden variable models considered and therefore on the definition of contextuality being considered.

In this example, we consider a simple additional restriction to hidden variable models compared to the traditional definition of non-contextuality: not all measurements within a single context may simultaneously take the value one. This restriction is purely artificial, but it is enough for the purpose of this example.

We can implement this constraint by defining a new class \texttt{RestrictedMeasurementScenario} that inherits from \texttt{MeasurementScenario} and modifies the \verb|incidence_matrix()| property. The direct consequence is that the computation of the contextual fraction and signalling fraction changes, and gives different results, as shown in Listing~\ref{lst:artifical_constraint}.

This example shows how the package can also be adapted to explore alternative notions of contextuality, different hidden variable models and in general provide flexibility, through modifications of the classes that implement measurement scenarios and empirical models.

\begin{listing}[!ht]
\begin{pythoncode}
from contextuality import EmpiricalModel, MeasurementScenario, MeasurementScenarioImplementations
import numpy as np

class RestrictedMeasurementScenario(MeasurementScenario):
    @property
    def incidence_matrix(self):
        incidence_matrix = super().incidence_matrix

        nb_contexts = len(self.M)
        nb_outcomes = len(self.all_outcomes)

        matrix = incidence_matrix.T.reshape(
            -1,
            nb_contexts,
            nb_outcomes,
        )

        impossibility_column = matrix[
            :,:,-1
        ]

        valid = ~np.any(impossibility_column > 0, axis=1)

        return incidence_matrix[:, valid]


ms = MeasurementScenarioImplementations.chsh()
restricted_ms = RestrictedMeasurementScenario(ms.X, ms.M, ms.O)

em = EmpiricalModel(ms, [[1,0,0,0], [0,1,0,0], [0,0,1,0], [0,0,0,1]])
em_restricted = EmpiricalModel(restricted_ms, em.vector)

cf = em.compute_cf()["CF"]
cf_restricted = em_restricted.compute_cf()["CF"]

print(f"CF: {round(cf, 4)}")
# OUT: 'CF: 0.0'
print(f"SF: {round(cf_restricted, 4)}")
# OUT: 'CF: 1.0'
\end{pythoncode}
\cprotect\caption{Listing to modify the incidence matrix so that it fits a new constraint. This is done through the implementation of a new class inheriting from the \verb|MeasurementScenario| class, therefore changing the value of the contextual fraction and signalling fraction.}
\label{lst:artifical_constraint}
\end{listing}

	\section{Benchmarks} \label{sec:benchmarks}

We present in this section the benchmark of two of the main methods of this package: the \verb|compute_cf()| and \verb|compute_sf()| methods of the \verb|EmpiricalModel| class.
Internally, these methods use linear programming, and hence the results depend on the solver used. Here the benchmarks are performed with two different solvers, MOSEK~\cite{mosek} and HiGHS~\cite{schwendinger2025HighsHiGHSOptimization} and with two different measurement scenarios: n-cycle scenarios~\cite{araujo2013AllNoncontextualityInequalities} and random scenarios which we detail in the following sections. As the package is not focused on optimization, these benchmark results are purely informative, to show that for a reasonably sized scenario, the methods are relatively fast. However, future developments might focus on improving performances.

For each benchmark point, the method was executed 100 times (\verb|compute_cf()| or \verb|compute_sf()|), excluding the time required for the creation of the measurement scenario. At each iteration, a new measurement scenario and a new random empirical model are instantiated. A random empirical model is produced in the following way, first we sample from a uniform distribution for each context, and then we normalize the sample such that it sums to 1. The code and the data used for these benchmarks are available in the \verb|benchmarks| directory on GitHub~\cite{vallee2025KimValleeContextualityGithub}.

\subsection{N-cycle scenarios} \label{ssec:n-cycle_scenarios}

For the first benchmarks, we consider a specific type of measurement scenarios called n-cycle measurement scenarios~\cite{araujo2013AllNoncontextualityInequalities}:

\begin{definition}[N-cycle measurement scenarios] \label{def:n-cycle_measurement_scenarios}%
An n-cycle measurement scenario is defined by a number $n$ with:
\begin{itemize}
	\item $X = \left\{A_0, \ldots, A_{n-1}\right\}$ the set of measurements such that $\left\lvert X \right\rvert = n$;
	\item $\mathcal{M} = \left\{A_0 A_1, A_1 A_2, \ldots, A_{n-2} A_{n-1}, A_{n-1} A_0\right\}$ the set of contexts;
	\item and $\forall x \in X\colon O_x = \left\{0,1\right\}$ the set of outcomes.
\end{itemize}
\end{definition}

N-cycle scenarios are convenient as they possess a well-defined structure. This means that constructing an arbitrary n-cycle scenario in the \contextuality{} package is straightforward and only requires knowing $n$, i.e., the size of the n-cycle scenario. 
Moreover, n-cycle scenarios are known to exhibit a gap between classical and quantum correlations~\cite{araujo2013AllNoncontextualityInequalities}. As a result, random empirical models can have a non-zero contextual fraction and therefore provide non-trivial benchmark instances. Overall the structure of n-cycle scenarios makes them very practical for benchmarking.

\begin{figure}[bth]
	\centering
	\begin{subfigure}[t]{0.50\linewidth}
		\centering%
		\includegraphics[width=\linewidth]{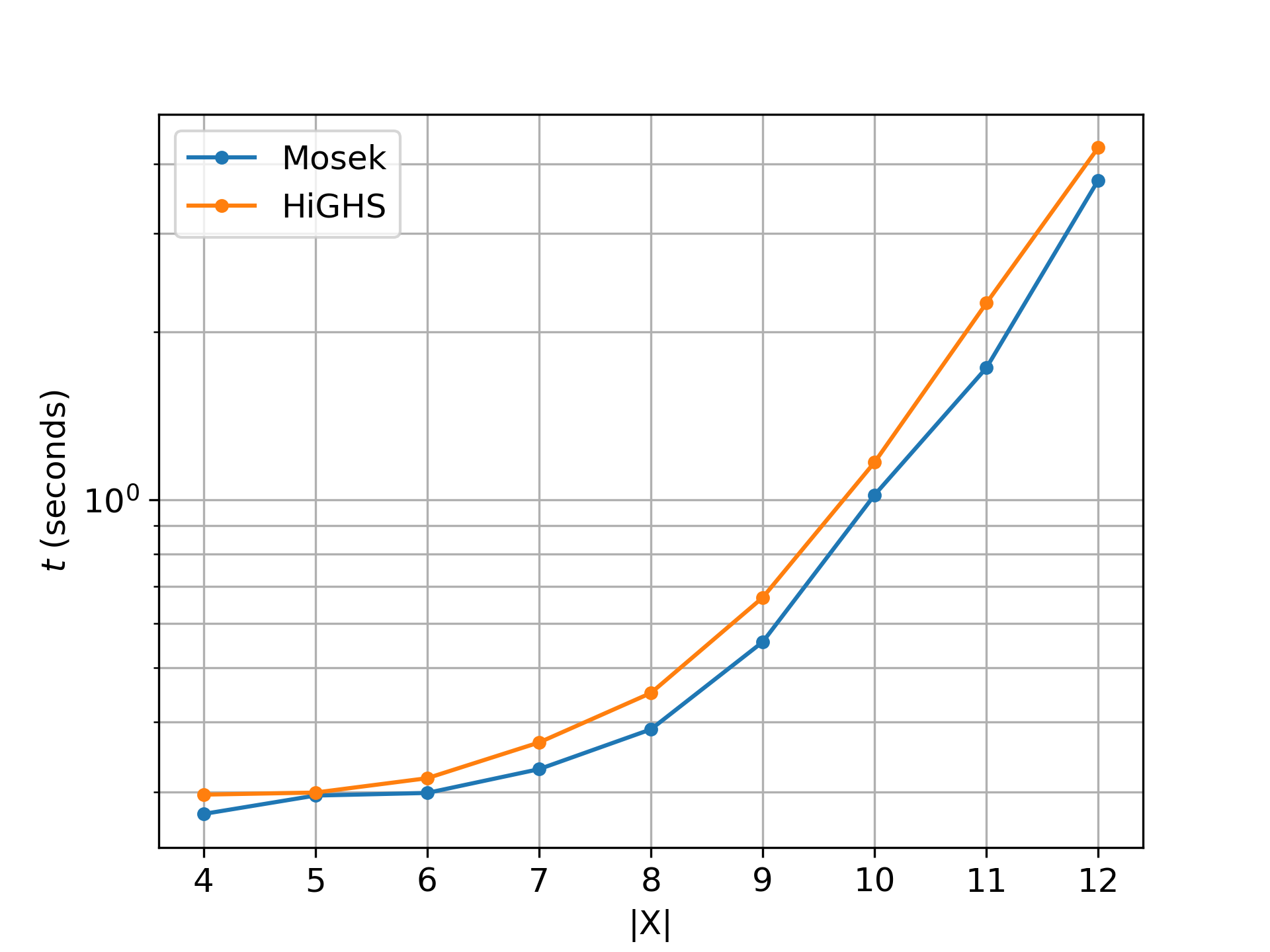}%
		\cprotect\caption{ \verb|compute_cf()|}%
		\label{fig:CF-n-cycle-benchmark}%
	\end{subfigure}%
	\begin{subfigure}[t]{0.50\linewidth}
		\centering%
		\includegraphics[width=\linewidth]{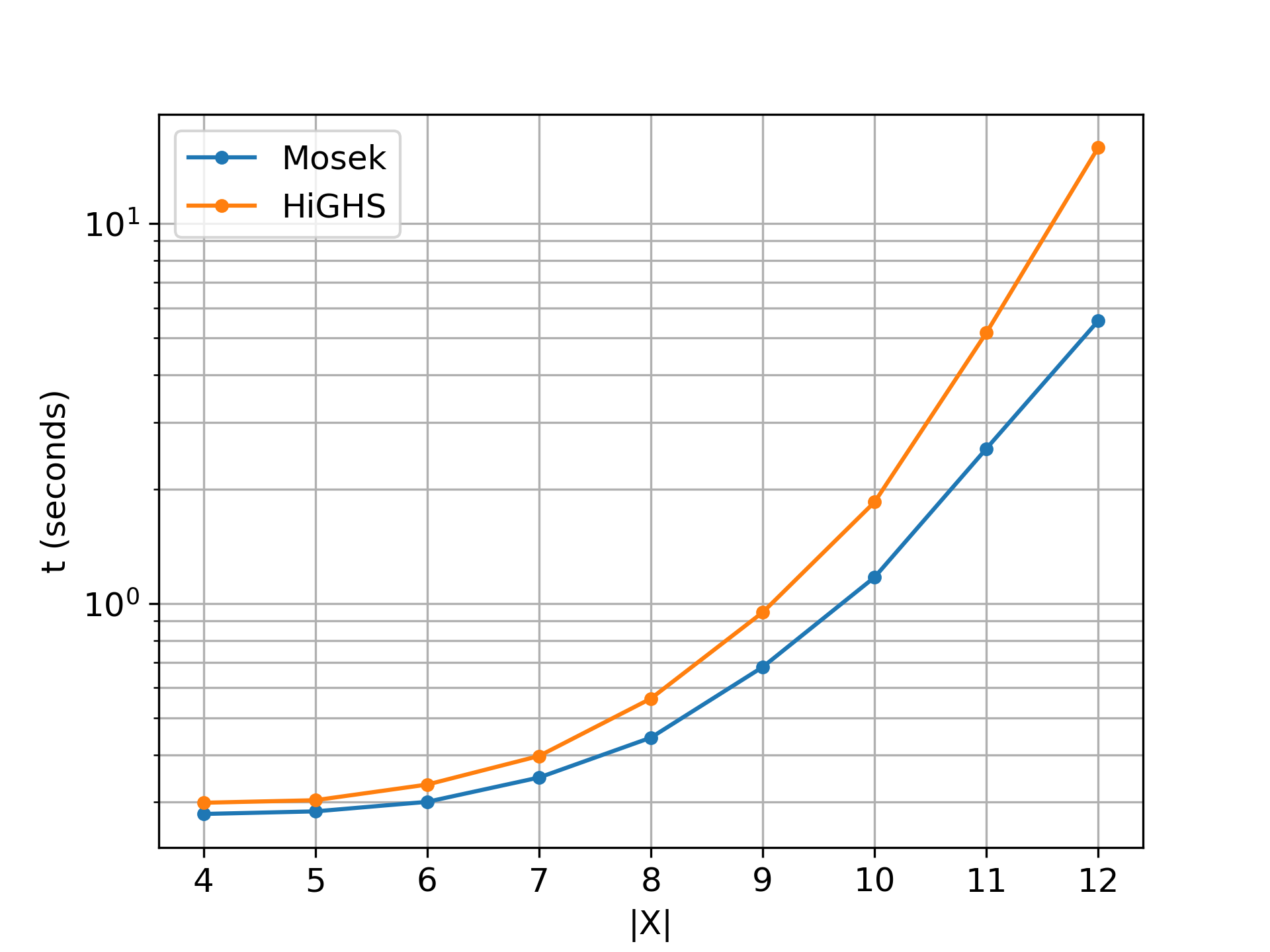}%
		\cprotect\caption{ \verb|compute_sf()|}%
		\label{fig:SF-n-cycle-benchmark}%
	\end{subfigure}
	\cprotect\caption{Benchmarks for n-cycle scenarios of increasing size $n$ for the \subref{fig:CF-n-cycle-benchmark} \verb|compute_cf()| method and for the \subref{fig:SF-n-cycle-benchmark} \verb|compute_sf()| method.
	The y-axis of each graph is in log-scale, therefore we see that we have at least an exponential complexity in time depending on the size of the measurement scenario.}
	\label{fig:benchmarks-n-cycle}
\end{figure}

The results of our benchmarks are presented in Figures~\ref{fig:CF-n-cycle-benchmark} and~\ref{fig:SF-n-cycle-benchmark}.

\subsection{Random scenarios} \label{ssec:random_scenarios}

Random scenarios are measurement scenarios specifically designed for these benchmarks. 
The generation of a random scenario of size $n$ essentially follows three steps:
\begin{enumerate}
	\item select a context size $K$ between $2$ and $\left\lfloor \frac{n}{2} \right\rfloor$, and a number of contexts $N$ proportional to  $\left\lfloor \frac{n}{K} \right\rfloor$;
	\item randomly sample $K$ measurements and add them as a context to the set of contexts. Repeat that step $N$ times;
	\item if some measurements were not added to any context, add them to more contexts.
\end{enumerate}
This procedure is made in order to avoid, as far as possible, trivial scenarios\footnote{Trivial measurement scenarios are those for which there is no difference between the classical, quantum and disturbing correlations.} and the full Python code can be found in the \verb|benchmarks| directory of the GitHub repository~\cite{vallee2025KimValleeContextualityGithub}.

\begin{figure}[htpb]
	\centering
	\begin{subfigure}[b]{0.5\textwidth}
		\centering
		\includegraphics[width=\linewidth]{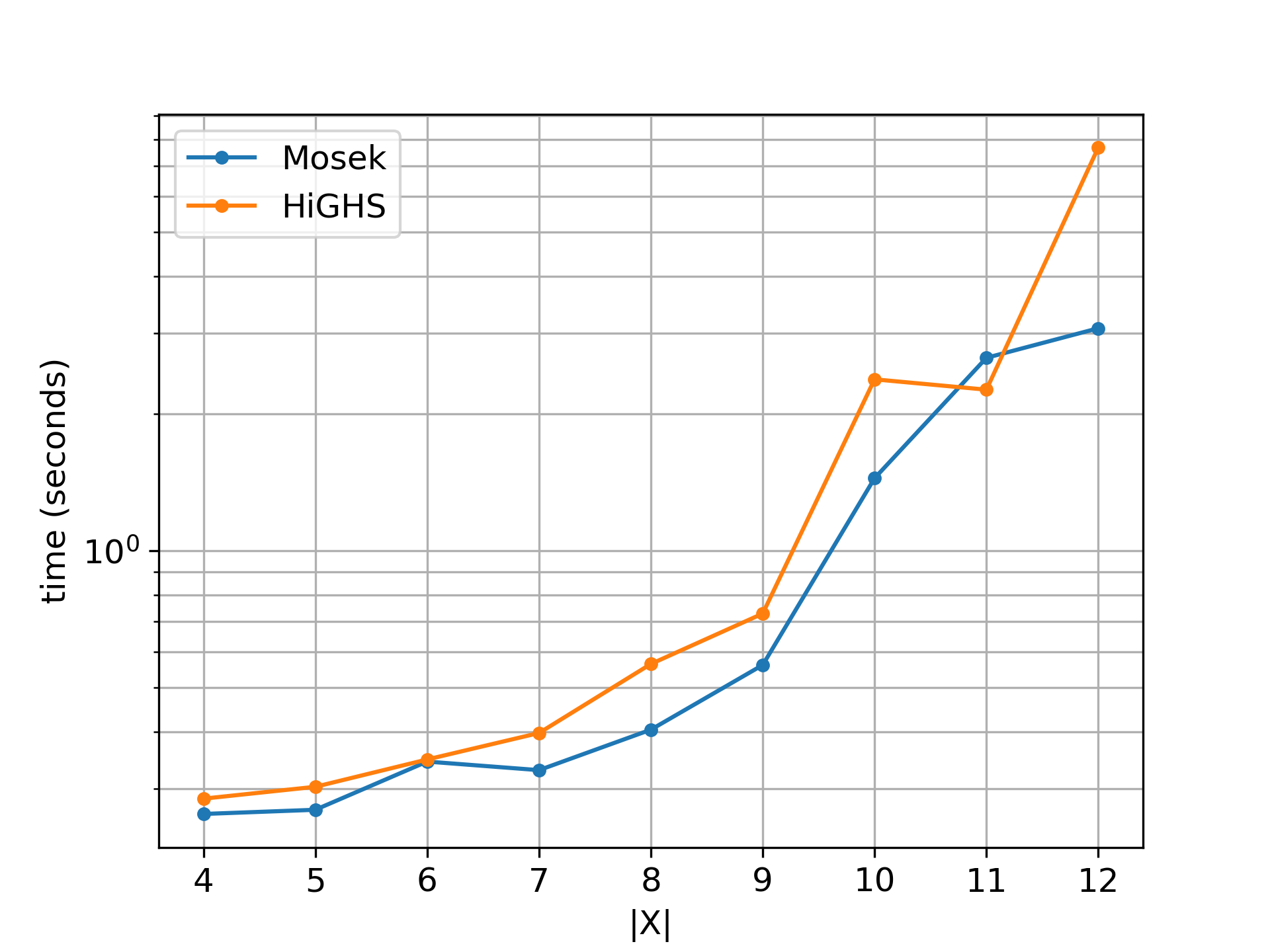}
		\cprotect\caption{\verb|compute_cf()|}
		\label{fig:cf-random-benchmark}
	\end{subfigure}%
	\begin{subfigure}[b]{0.5\textwidth}
		\centering
		\includegraphics[width=\linewidth]{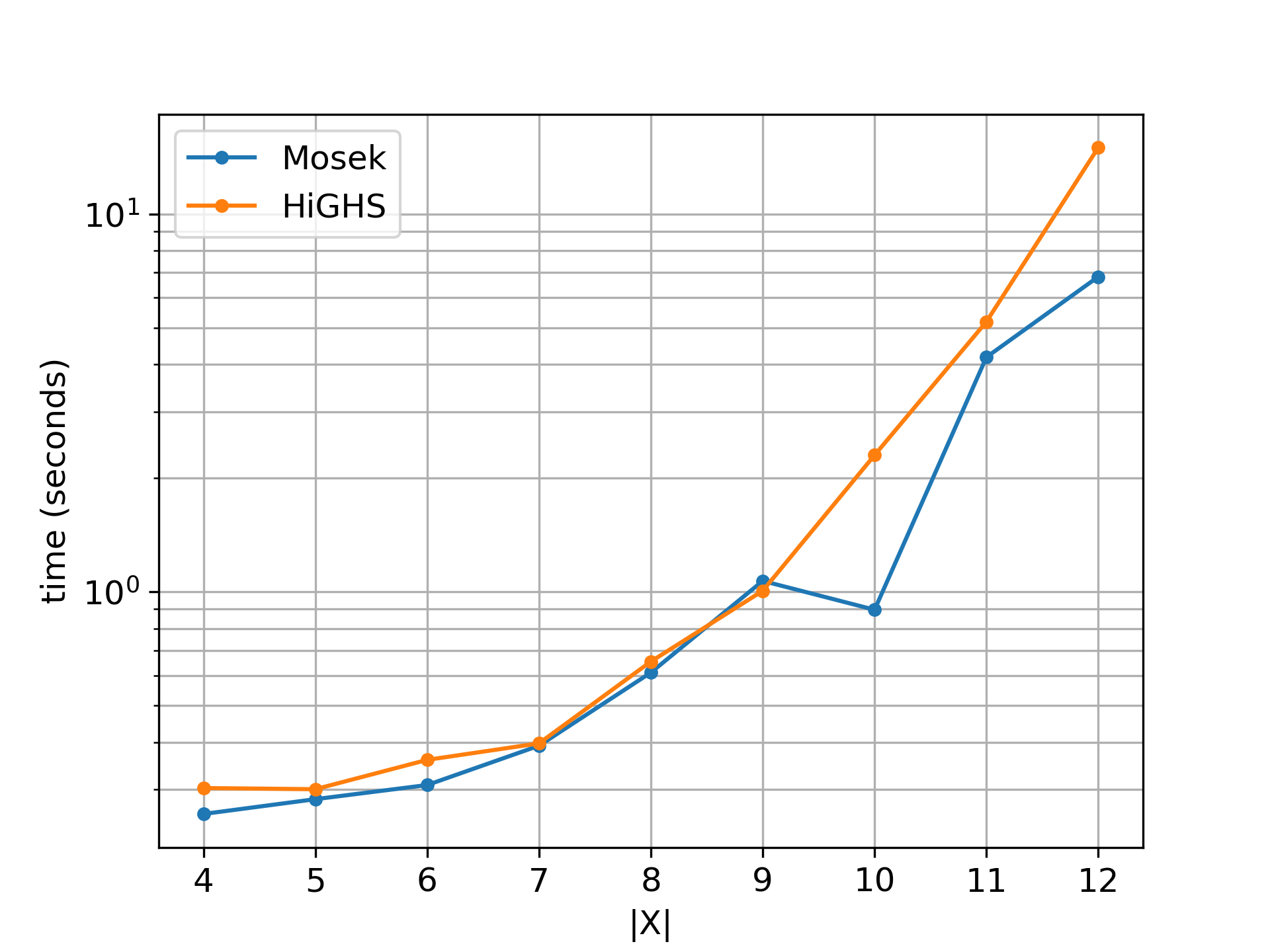}
		\cprotect\caption{\verb|compute_sf()|}
		\label{fig:sf-random-benchmark}
	\end{subfigure}
	\cprotect\caption{Benchmarks for random scenarios of increasing size $n$ for the \subref{fig:CF-n-cycle-benchmark} \verb|compute_cf()| method and for the \subref{fig:SF-n-cycle-benchmark} \verb|compute_sf()| method.
	The y-axis of each graph is in log-scale, however the complexity is more difficult to analyse due to the random nature of the scenarios.}
	\label{fig:random-benchmarks}
\end{figure}

Contrary to n-cycle scenarios, random scenarios do not have a well-defined structure, and are not certified to be non-trivial despite the procedure. This leads to more variable performances as we can see in the benchmarks reported in Figures~\ref{fig:cf-random-benchmark} and~\ref{fig:sf-random-benchmark}.

	\section{Conclusion} \label{sec:discussion}

In this paper we presented the \contextuality{} package which contains many features related to the sheaf-theoretic approach to contextuality~\cite{abramsky2011SheaftheoreticStructureNonlocality}. It provides a simplified way to construct and manipulate measurement scenarios and empirical models as shown in Section~\ref{sec:core_features}. In these sections we provided small pedagogical examples on how to use the package. We completed this section with the presentation of the miscellaneous module \texttt{utils} which provides a lot of experimental and legacy functions.

Additionally, we also gave more in-depth examples of the package's use for two experimental papers~\cite{lapkiewicz2011ExperimentalNonclassicalityIndivisible,wang2022SignificantLoopholefreeTest} in Section~\ref{sec:example_usage}. 
These examples illustrate the simplicity of the package and provide useful insights into its use.

Together, these examples and benchmarks show how \contextuality{} can support both the exploration of standard contextuality scenarios and the development of more specialized cases.

\subsection{Future developments} \label{ssec:future_developments}

The main goal of this project is to provide a Python package that can be used by many scientists in quantum foundations. In the short term, we wish to improve the documentation, add more practical methods, for instance to simplify the creation of standard empirical models or measurement scenarios, and finally work on the dissemination of the package.
In the long term, we believe it can be extended to include other frameworks of contextuality, such as generalized contextuality~\cite{spekkens2005ContextualityPreparationsTransformations,schmid2018AllNoncontextualityInequalities}, the graph theoretic approach~\cite{cabello2014GraphTheoreticApproachQuantum}, or the hypergraph approach~\cite{acin2015CombinatorialApproachNonlocality}. 
Readers interested in implementing these frameworks are encouraged to contact the author or submit pull requests directly.

	\section*{Acknowledgments} 

	We acknowledge the help of Yoann Piétri, for his thorough proofread and inputs, Adel Sohbi with whom we started developing this package and of Damian Markham, whose discussions led to improvements of the package. We also thank Dana Evans for her valuable input on the manuscript.
	We acknowledge financial support from the French national quantum initiative managed by Agence Nationale de la Recherche in the framework of France 2030 through project EPIQ, with the reference ANR-22-PETQ-0007.

	\printbibliography

@article{abramsky2011SheaftheoreticStructureNonlocality,
  title = {The Sheaf-Theoretic Structure of Non-Locality and Contextuality},
  author = {Abramsky, Samson and Brandenburger, Adam},
  date = {2011-11},
  journaltitle = {New Journal of Physics},
  shortjournal = {New J. Phys.},
  volume = {13},
  number = {11},
  pages = {113036},
  publisher = {IOP Publishing},
  issn = {1367-2630},
  doi = {10.1088/1367-2630/13/11/113036},
  langid = {english}
}

@article{abramsky2017ContextualFractionMeasure,
  title = {The Contextual Fraction as a Measure of Contextuality},
  author = {Abramsky, Samson and Barbosa, Rui Soares and Mansfield, Shane},
  date = {2017-08-04},
  journaltitle = {Physical Review Letters},
  shortjournal = {Phys. Rev. Lett.},
  volume = {119},
  number = {5},
  eprint = {1705.07918},
  eprinttype = {arXiv},
  pages = {050504},
  issn = {0031-9007, 1079-7114},
  doi = {10.1103/PhysRevLett.119.050504}
}

@article{acin2015CombinatorialApproachNonlocality,
  title = {A {{Combinatorial Approach}} to {{Nonlocality}} and {{Contextuality}}},
  author = {Ac\'in, Antonio and Fritz, Tobias and Leverrier, Anthony and Sainz, Ana Bel\'en},
  date = {2015-03-01},
  journaltitle = {Communications in Mathematical Physics},
  shortjournal = {Commun. Math. Phys.},
  volume = {334},
  number = {2},
  pages = {533--628},
  issn = {1432-0916},
  doi = {10.1007/s00220-014-2260-1},
  langid = {english}
}

@article{araujo2013AllNoncontextualityInequalities,
  title = {All Noncontextuality Inequalities for the N-Cycle Scenario},
  author = {Ara\'ujo, Mateus and Quintino, Marco T\'ulio and Budroni, Costantino and Cunha, Marcelo Terra and Cabello, Ad\'an},
  date = {2013-08-21},
  journaltitle = {Physical Review A},
  shortjournal = {Phys. Rev. A},
  volume = {88},
  number = {2},
  pages = {022118},
  publisher = {American Physical Society},
  doi = {10.1103/PhysRevA.88.022118}
}

@software{araujo2025KetjlJuliaToolbox,
  title = {Ket.Jl: A {{Julia}} Toolbox for Quantum Information, Nonlocality, and Entanglement},
  shorttitle = {Ket.Jl},
  author = {Ara\'ujo, Mateus and Brown, Peter and Designolle, S\'ebastien and family=Gois, given=Carlos, prefix=de, useprefix=true and Liu, Ye-Chao and Porto, Lucas and Quintino, Marco T\'ulio},
  date = {2025-11-10},
  doi = {10.5281/ZENODO.17574731},
  organization = {Zenodo},
  version = {v0.8.0}
}

@article{bell1964EinsteinPodolskyRosen,
  title = {On the {{Einstein Podolsky Rosen}} Paradox},
  author = {Bell, J. S.},
  date = {1964-11-01},
  journaltitle = {Physics Physique Fizika},
  shortjournal = {Phys. Phys. Fiz.},
  volume = {1},
  number = {3},
  pages = {195--200},
  publisher = {American Physical Society},
  doi = {10.1103/PhysicsPhysiqueFizika.1.195}
}

@article{bermejo-vega2017ContextualityResourceModels,
  title = {Contextuality as a {{Resource}} for {{Models}} of {{Quantum Computation}} with {{Qubits}}},
  author = {Bermejo-Vega, Juan and Delfosse, Nicolas and Browne, Dan E. and Okay, Cihan and Raussendorf, Robert},
  date = {2017-09-21},
  journaltitle = {Physical Review Letters},
  shortjournal = {Phys. Rev. Lett.},
  volume = {119},
  number = {12},
  pages = {120505},
  publisher = {American Physical Society},
  doi = {10.1103/PhysRevLett.119.120505}
}

@article{budroni2022KochenSpeckerContextuality,
  title = {Kochen-{{Specker}} Contextuality},
  author = {Budroni, Costantino and Cabello, Ad\'an and G\"uhne, Otfried and Kleinmann, Matthias and Larsson, Jan-\AA ke},
  date = {2022-12-19},
  journaltitle = {Reviews of Modern Physics},
  shortjournal = {Rev. Mod. Phys.},
  volume = {94},
  number = {4},
  pages = {045007},
  publisher = {American Physical Society},
  doi = {10.1103/RevModPhys.94.045007},
  langid = {english}
}

@article{cabello2008ExperimentallyTestableStateIndependent,
  title = {Experimentally {{Testable State-Independent Quantum Contextuality}}},
  author = {Cabello, Ad\'an},
  date = {2008-11-19},
  journaltitle = {Physical Review Letters},
  shortjournal = {Phys. Rev. Lett.},
  volume = {101},
  number = {21},
  pages = {210401},
  publisher = {American Physical Society},
  doi = {10.1103/PhysRevLett.101.210401}
}

@online{cabello2010NonContextualityPhysicalTheories,
  title = {({{Non-}}){{Contextuality}} of {{Physical Theories}} as an {{Axiom}}},
  author = {Cabello, Adan and Severini, Simone and Winter, Andreas},
  date = {2010-10-11},
  eprint = {1010.2163},
  eprinttype = {arXiv},
  eprintclass = {quant-ph},
  doi = {10.48550/arXiv.1010.2163},
  pubstate = {prepublished}
}

@article{cabello2014GraphTheoreticApproachQuantum,
  title = {Graph-{{Theoretic Approach}} to {{Quantum Correlations}}},
  author = {Cabello, Ad\'an and Severini, Simone and Winter, Andreas},
  date = {2014-01-27},
  journaltitle = {Physical Review Letters},
  shortjournal = {Phys. Rev. Lett.},
  volume = {112},
  number = {4},
  pages = {040401},
  publisher = {American Physical Society},
  doi = {10.1103/PhysRevLett.112.040401}
}

@article{clauser1969ProposedExperimentTest,
  title = {Proposed {{Experiment}} to {{Test Local Hidden-Variable Theories}}},
  author = {Clauser, John F. and Horne, Michael A. and Shimony, Abner and Holt, Richard A.},
  date = {1969-10-13},
  journaltitle = {Physical Review Letters},
  shortjournal = {Phys. Rev. Lett.},
  volume = {23},
  number = {15},
  pages = {880--884},
  publisher = {American Physical Society},
  doi = {10.1103/PhysRevLett.23.880}
}

@online{dzhafarov2016ContextualitybyDefault20Systems,
  title = {Contextuality-by-{{Default}} 2.0: {{Systems}} with {{Binary Random Variables}}},
  shorttitle = {Contextuality-by-{{Default}} 2.0},
  author = {Dzhafarov, Ehtibar N. and Kujala, Janne V.},
  date = {2016-10-25},
  eprint = {1604.04799},
  eprinttype = {arXiv},
  eprintclass = {quant-ph},
  doi = {10.48550/arXiv.1604.04799},
  pubstate = {prepublished}
}

@online{GettingStartedPycddlib,
  title = {Getting {{Started}} --- Pycddlib 3.0.2 Documentation},
  url = {https://pycddlib.readthedocs.io/en/stable/quickstart.html#installing-cddlib-and-gmp},
  urldate = {2026-03-09}
}

@misc{gplv3,
  title = {{{GNU}} General Public License},
  shorttitle = {{{GNU GPLv3}}},
  date = {2007-06-29},
  url = {https://www.gnu.org/licenses/gpl-3.0.html},
  urldate = {2026-07-15},
  langid = {english},
  organization = {Free Software Foundation},
  pagination = {section},
  version = {3}
}

@article{howard2014ContextualitySuppliesMagic,
  title = {Contextuality Supplies the `Magic' for Quantum Computation},
  author = {Howard, Mark and Wallman, Joel and Veitch, Victor and Emerson, Joseph},
  date = {2014-06},
  journaltitle = {Nature},
  volume = {510},
  number = {7505},
  pages = {351--355},
  publisher = {Nature Publishing Group},
  issn = {1476-4687},
  doi = {10.1038/nature13460},
  issue = {7505},
  langid = {english}
}

@article{klyachko2008SimpleTestHidden,
  title = {Simple {{Test}} for {{Hidden Variables}} in {{Spin-1 Systems}}},
  author = {Klyachko, Alexander A. and Can, M. Ali and Binicio\u glu, Sinem and Shumovsky, Alexander S.},
  date = {2008-07-11},
  journaltitle = {Physical Review Letters},
  shortjournal = {Phys. Rev. Lett.},
  volume = {101},
  number = {2},
  pages = {020403},
  publisher = {American Physical Society},
  doi = {10.1103/PhysRevLett.101.020403}
}

@article{kochen1967ProblemHiddenVariables,
  title = {The {{Problem}} of {{Hidden Variables}} in {{Quantum Mechanics}}},
  author = {Kochen, Simon and Specker, E. P.},
  date = {1967},
  journaltitle = {Journal of Mathematics and Mechanics},
  shortjournal = {J. Math. Mech.},
  volume = {17},
  number = {1},
  pages = {59--87},
  publisher = {Indiana University Mathematics Department},
  issn = {0095-9057},
  url = {https://www.jstor.org/stable/24902153},
  urldate = {2022-01-11}
}

@article{krishna2017DerivingRobustNoncontextuality,
  title = {Deriving Robust Noncontextuality Inequalities from Algebraic Proofs of the {{Kochen-Specker}} Theorem: The {{Peres-Mermin}} Square},
  shorttitle = {Deriving Robust Noncontextuality Inequalities from Algebraic Proofs of the {{Kochen-Specker}} Theorem},
  author = {Krishna, Anirudh and Spekkens, Robert W. and Wolfe, Elie},
  date = {2017-12-14},
  journaltitle = {New Journal of Physics},
  shortjournal = {New J. Phys.},
  volume = {19},
  number = {12},
  eprint = {1704.01153},
  eprinttype = {arXiv},
  eprintclass = {quant-ph},
  pages = {123031},
  issn = {1367-2630},
  doi = {10.1088/1367-2630/aa9168}
}

@online{lambert2025QuTiP5Quantum,
  title = {{{QuTiP}} 5: {{The Quantum Toolbox}} in {{Python}}},
  shorttitle = {{{QuTiP}} 5},
  author = {Lambert, Neill and Gigu\`ere, Eric and Menczel, Paul and Li, Boxi and Hopf, Patrick and Su\'arez, Gerardo and Gali, Marc and Lishman, Jake and Gadhvi, Rushiraj and Agarwal, Rochisha and Galicia, Asier and Shammah, Nathan and Nation, Paul and Johansson, J. R. and Ahmed, Shahnawaz and Cross, Simon and Pitchford, Alexander and Nori, Franco},
  date = {2025-10-01},
  eprint = {2412.04705},
  eprinttype = {arXiv},
  eprintclass = {quant-ph},
  doi = {10.48550/arXiv.2412.04705},
  pubstate = {prepublished}
}

@article{lapkiewicz2011ExperimentalNonclassicalityIndivisible,
  title = {Experimental Non-Classicality of an Indivisible Quantum System},
  author = {Lapkiewicz, Radek and Li, Peizhe and Schaeff, Christoph and Langford, Nathan K. and Ramelow, Sven and Wie\'sniak, Marcin and Zeilinger, Anton},
  date = {2011-06},
  journaltitle = {Nature},
  volume = {474},
  number = {7352},
  pages = {490--493},
  publisher = {Nature Publishing Group},
  issn = {1476-4687},
  doi = {10.1038/nature10119},
  issue = {7352},
  langid = {english}
}

@article{mazurek2016ExperimentalTestNoncontextuality,
  title = {An Experimental Test of Noncontextuality without Unphysical Idealizations},
  author = {Mazurek, Michael D. and Pusey, Matthew F. and Kunjwal, Ravi and Resch, Kevin J. and Spekkens, Robert W.},
  date = {2016-06-13},
  journaltitle = {Nature Communications},
  shortjournal = {Nat. Commun.},
  volume = {7},
  number = {1},
  pages = {ncomms11780},
  publisher = {Nature Publishing Group},
  issn = {2041-1723},
  doi = {10.1038/ncomms11780},
  issue = {1},
  langid = {english}
}

@article{mercurio2025QuantumToolboxjlEfficientJulia,
  title = {{{QuantumToolbox}}.Jl: {{An}} Efficient {{Julia}} Framework for Simulating Open Quantum Systems},
  shorttitle = {{{QuantumToolbox}}.Jl},
  author = {Mercurio, Alberto and Huang, Yi-Te and Cai, Li-Xun and Chen, Yueh-Nan and Savona, Vincenzo and Nori, Franco},
  date = {2025-09-29},
  journaltitle = {Quantum},
  volume = {9},
  pages = {1866},
  publisher = {Verein zur F\"orderung des Open Access Publizierens in den Quantenwissenschaften},
  doi = {10.22331/q-2025-09-29-1866},
  langid = {british}
}

@article{mermin1990SimpleUnifiedForm,
  title = {Simple Unified Form for the Major No-Hidden-Variables Theorems},
  author = {Mermin, N. David},
  date = {1990-12-31},
  journaltitle = {Physical Review Letters},
  shortjournal = {Phys. Rev. Lett.},
  volume = {65},
  number = {27},
  pages = {3373--3376},
  publisher = {American Physical Society},
  doi = {10.1103/PhysRevLett.65.3373}
}

@article{meurer2017SymPySymbolicComputing,
  title = {{{SymPy}}: Symbolic Computing in {{Python}}},
  shorttitle = {{{SymPy}}},
  author = {Meurer, Aaron and Smith, Christopher P. and Paprocki, Mateusz and \v Cert\'ik, Ond\v rej and Kirpichev, Sergey B. and Rocklin, Matthew and family=Kumar, given=AmiT, given-i={{Am}} and Ivanov, Sergiu and Moore, Jason K. and Singh, Sartaj and Rathnayake, Thilina and Vig, Sean and Granger, Brian E. and Muller, Richard P. and Bonazzi, Francesco and Gupta, Harsh and Vats, Shivam and Johansson, Fredrik and Pedregosa, Fabian and Curry, Matthew J. and Terrel, Andy R. and Rou\v cka, \v St\v ep\'an and Saboo, Ashutosh and Fernando, Isuru and Kulal, Sumith and Cimrman, Robert and Scopatz, Anthony},
  date = {2017-01-02},
  journaltitle = {PeerJ Computer Science},
  shortjournal = {PeerJ Comput. Sci.},
  volume = {3},
  pages = {e103},
  issn = {2376-5992},
  doi = {10.7717/peerj-cs.103},
  langid = {english}
}

@manual{mosek,
  type = {manual},
  title = {{{MOSEK}} Optimizer {{API}} for Python 11.0.29},
  author = {family=ApS, given=MOSEK, given-i=MOSEK},
  date = {2019},
  url = {https://docs.mosek.com/11.0/pythonapi/index.html}
}

@article{peres1991TwoSimpleProofs,
  title = {Two Simple Proofs of the {{Kochen-Specker}} Theorem},
  author = {Peres, A.},
  date = {1991-02},
  journaltitle = {Journal of Physics A: Mathematical and General},
  shortjournal = {J. Phys. Math. Gen.},
  volume = {24},
  number = {4},
  pages = {L175--L178},
  publisher = {IOP Publishing},
  issn = {0305-4470},
  doi = {10.1088/0305-4470/24/4/003},
  langid = {english}
}

@article{popescu1994QuantumNonlocalityAxiom,
  title = {Quantum Nonlocality as an Axiom},
  author = {Popescu, Sandu and Rohrlich, Daniel},
  date = {1994-03-01},
  journaltitle = {Foundations of Physics},
  shortjournal = {Found. Phys.},
  volume = {24},
  number = {3},
  pages = {379--385},
  issn = {1572-9516},
  doi = {10.1007/BF02058098},
  langid = {english}
}

@article{raussendorf2013ContextualityMeasurementbasedQuantum,
  title = {Contextuality in Measurement-Based Quantum Computation},
  author = {Raussendorf, Robert},
  date = {2013-08-19},
  journaltitle = {Physical Review A},
  shortjournal = {Phys. Rev. A},
  volume = {88},
  number = {2},
  pages = {022322},
  publisher = {American Physical Society},
  doi = {10.1103/PhysRevA.88.022322}
}

@software{russo2020ToqitoTheoryQuantum,
  title = {Toqito -- {{Theory}} of Quantum Information Toolkit: {{A Python}} Package for Studying Quantum Information},
  shorttitle = {Toqito -- {{Theory}} of Quantum Information Toolkit},
  author = {Russo, Vincent},
  date = {2020-03-06},
  doi = {10.5281/ZENODO.4743211},
  langid = {english},
  organization = {Zenodo},
  version = {v1.0}
}

@article{schmid2018AllNoncontextualityInequalities,
  title = {All the Noncontextuality Inequalities for Arbitrary Prepare-and-Measure Experiments with Respect to Any Fixed Set of Operational Equivalences},
  author = {Schmid, David and Spekkens, Robert W. and Wolfe, Elie},
  date = {2018-06-05},
  journaltitle = {Physical Review A},
  shortjournal = {Phys. Rev. A},
  volume = {97},
  number = {6},
  pages = {062103},
  publisher = {American Physical Society},
  doi = {10.1103/PhysRevA.97.062103}
}

@article{schmid2018ContextualAdvantageState,
  title = {Contextual Advantage for State Discrimination},
  author = {Schmid, David and Spekkens, Robert W.},
  date = {2018-02-02},
  journaltitle = {Physical Review X},
  shortjournal = {Phys. Rev. X},
  volume = {8},
  number = {1},
  eprint = {1706.04588},
  eprinttype = {arXiv},
  eprintclass = {quant-ph},
  pages = {011015},
  issn = {2160-3308},
  doi = {10.1103/PhysRevX.8.011015}
}

@software{schwendinger2025HighsHiGHSOptimization,
  title = {Highs: '{{HiGHS}}' {{Optimization Solver}}},
  shorttitle = {Highs},
  author = {Schwendinger, Florian and Schumacher, Dirk and Hall, Julian and Galabova, Ivet and Gottwald, Leona and Feldmeier, Michael},
  date = {2025-07-13},
  url = {https://cran.r-project.org/web/packages/highs/index.html},
  urldate = {2025-10-31},
  version = {1.10.0-3}
}

@article{selby2022OpensourceLinearProgram,
  title = {An Open-Source Linear Program for Testing Nonclassicality},
  author = {Selby, John H. and Wolfe, Elie and Schmid, David and Sainz, Ana Bel\'en},
  date = {2022-04-25},
  journaltitle = {ArXiv220411905 Quant-Ph},
  eprint = {2204.11905},
  eprinttype = {arXiv},
  eprintclass = {quant-ph},
  doi = {10.48550/arXiv.2204.11905}
}

@book{skrzypczyk2023SemidefiniteProgrammingQuantum,
  title = {Semidefinite {{Programming}} in {{Quantum Information Science}}},
  author = {Skrzypczyk, Paul and Cavalcanti, Daniel},
  date = {2023-03-01},
  publisher = {IOP Publishing},
  url = {https://iopscience.iop.org/book/mono/978-0-7503-3343-6},
  urldate = {2023-09-15},
  isbn = {978-0-7503-3343-6},
  langid = {english}
}

@article{spekkens2005ContextualityPreparationsTransformations,
  title = {Contextuality for Preparations, Transformations, and Unsharp Measurements},
  author = {Spekkens, R. W.},
  date = {2005-05-31},
  journaltitle = {Physical Review A},
  shortjournal = {Phys. Rev. A},
  volume = {71},
  number = {5},
  pages = {052108},
  publisher = {American Physical Society},
  doi = {10.1103/PhysRevA.71.052108}
}

@article{spekkens2009PreparationContextualityPowers,
  title = {Preparation {{Contextuality Powers Parity-Oblivious Multiplexing}}},
  author = {Spekkens, Robert W. and Buzacott, D. H. and Keehn, A. J. and Toner, Ben and Pryde, G. J.},
  date = {2009-01-05},
  journaltitle = {Physical Review Letters},
  shortjournal = {Phys. Rev. Lett.},
  volume = {102},
  number = {1},
  pages = {010401},
  publisher = {American Physical Society},
  doi = {10.1103/PhysRevLett.102.010401}
}

@article{steiger2018ProjectQOpenSource,
  title = {{{ProjectQ}}: An Open Source Software Framework for Quantum Computing},
  shorttitle = {{{ProjectQ}}},
  author = {Steiger, Damian S. and H\"aner, Thomas and Troyer, Matthias},
  date = {2018-01-31},
  journaltitle = {Quantum},
  volume = {2},
  pages = {49},
  publisher = {Verein zur F\"orderung des Open Access Publizierens in den Quantenwissenschaften},
  doi = {10.22331/q-2018-01-31-49},
  langid = {british}
}

@article{vallee2024CorrectedBellNoncontextuality,
  title = {Corrected {{Bell}} and Non-Contextuality Inequalities for Realistic Experiments},
  author = {Vall\'ee, Kim and Emeriau, Pierre-Emmanuel and Bourdoncle, Boris and Sohbi, Adel and Mansfield, Shane and Markham, Damian},
  date = {2024-01-29},
  journaltitle = {Philosophical Transactions of the Royal Society A: Mathematical, Physical and Engineering Sciences},
  shortjournal = {Philos. Trans. R. Soc. Math. Phys. Eng. Sci.},
  volume = {382},
  number = {2268},
  pages = {20230011},
  publisher = {Royal Society},
  doi = {10.1098/rsta.2023.0011}
}

@online{vallee2025ContextualityPackagePyPi,
  type = {Repository},
  title = {{{PyPi}} Repository: Contextuality},
  shorttitle = {Contextuality},
  author = {Vall\'ee, Kim},
  date = {2025},
  url = {https://pypi.org/project/contextuality/},
  urldate = {2025-10-29},
  organization = {PyPi}
}

@online{vallee2025KimValleeContextualityGithub,
  type = {Repository},
  title = {Kim-{{Vallee}}/Contextuality},
  author = {Vall\'ee, Kim},
  date = {2025},
  origdate = {2022-03-07},
  url = {https://github.com/Kim-Vallee/contextuality},
  urldate = {2025-10-29},
  organization = {GitHub}
}

@online{vallee2025RTDContextuality,
  type = {Documentation},
  title = {Contextuality Documentation},
  author = {Vall\'ee, Kim},
  date = {2025-10-29},
  url = {https://contextuality.readthedocs.io/en/latest/},
  urldate = {2025-10-29},
  organization = {ReadtheDocs}
}

@article{wang2022SignificantLoopholefreeTest,
  title = {Significant Loophole-Free Test of {{Kochen-Specker}} Contextuality Using Two Species of Atomic Ions},
  author = {Wang, Pengfei and Zhang, Junhua and Luan, Chun-Yang and Um, Mark and Wang, Ye and Qiao, Mu and Xie, Tian and Zhang, Jing-Ning and Cabello, Ad\'an and Kim, Kihwan},
  date = {2022-02-09},
  journaltitle = {Science Advances},
  shortjournal = {Sci. Adv.},
  volume = {8},
  number = {6},
  pages = {eabk1660},
  publisher = {American Association for the Advancement of Science},
  doi = {10.1126/sciadv.abk1660}
}

@article{yu2012StateIndependentProofKochenSpecker,
  title = {State-{{Independent Proof}} of {{Kochen-Specker Theorem}} with 13 {{Rays}}},
  author = {Yu, Sixia and Oh, C. H.},
  date = {2012-01-18},
  journaltitle = {Physical Review Letters},
  shortjournal = {Phys. Rev. Lett.},
  volume = {108},
  number = {3},
  pages = {030402},
  publisher = {American Physical Society},
  doi = {10.1103/PhysRevLett.108.030402}
}
	
	\clearpage

	\appendix
	
	\section{Sheaf theoretic framework for contextuality} \label{sec:sheaf_theoretic_framework_for_contextuality}

Since the development of quantum mechanics, scientists have sought to understand its features.
One of them is contextuality~\cite{kochen1967ProblemHiddenVariables}, which has attracted increasing interest in the past decades, from the theoretical point of view~\cite{spekkens2005ContextualityPreparationsTransformations,budroni2022KochenSpeckerContextuality,abramsky2011SheaftheoreticStructureNonlocality,dzhafarov2016ContextualitybyDefault20Systems,cabello2014GraphTheoreticApproachQuantum,acin2015CombinatorialApproachNonlocality} as well as for its potential applications~\cite{howard2014ContextualitySuppliesMagic,spekkens2009PreparationContextualityPowers,schmid2018ContextualAdvantageState,raussendorf2013ContextualityMeasurementbasedQuantum,bermejo-vega2017ContextualityResourceModels}.
	
	A common way to demonstrate contextuality is through the derivation of an appropriate inequality~\cite{kochen1967ProblemHiddenVariables,klyachko2008SimpleTestHidden,cabello2008ExperimentallyTestableStateIndependent,krishna2017DerivingRobustNoncontextuality,mazurek2016ExperimentalTestNoncontextuality,yu2012StateIndependentProofKochenSpecker}.
	Nevertheless, with each new scenario comes a new inequality -- or more generally multiple inequalities -- to check whether a behaviour is contextual and how contextual it is. Finding each inequality by hand is rather cumbersome, but the sheaf theoretic approach to contextuality~\cite{abramsky2011SheaftheoreticStructureNonlocality} solves this issue with a consistent framework and linear programming~\cite{abramsky2017ContextualFractionMeasure}.

\subsection{Overview of the sheaf theoretic framework for contextuality} \label{ssec:overview_of_sheaf_theoretic_contextuality}

In this section we give a very brief overview of the sheaf theoretic framework for contextuality and associated quantities, as familiarity with this framework is required to use the package. A more in-depth presentation can be found in the original paper~\cite{abramsky2011SheaftheoreticStructureNonlocality}.

In this framework, the most general abstraction of an experimental setup is described by a \emph{measurement scenario}:
\begin{definition}[Measurement scenario] \label{def:measurement_scenario}%
A measurement scenario is composed of a triple $\XMO$ where:
\begin{itemize}
	\item $X$ is a set of measurement labels;
	\item $\mathcal{M}$ is a set of subsets $C$ of $X$, where each $C \in \mathcal{M}$ is called a measurement context;
	\item $O$ is the set of outcomes for each measurement in $X$, and we write $O^{C}$ as the set of outcomes of any subset $C \subseteq X$.
\end{itemize}
\end{definition}

The most common example of a measurement scenario is the CHSH scenario~\cite{bell1964EinsteinPodolskyRosen,clauser1969ProposedExperimentTest} described by the triple $\XMO[\text{CHSH}]$:
\begin{subequations} \label{eqs:CHSH_scenario}
	\begin{align}
		X_{\text{CHSH}} &= \left\{A, A', B, B'\right\} \\
		\mathcal{M}_{\text{CHSH}} &= \left\{ \left\{ A, B\right\}, \left\{A, B'\right\}, \left\{A', B\right\}, \left\{A', B'\right\} \right\} \\
		O_{\text{CHSH}} &= \left\{0, 1\right\}
	.\end{align}
\end{subequations}

A realization of a measurement scenario is a set of probability distributions obtained either from a theoretical model or from experimental data. Such a realization is called an \emph{empirical model}:
\begin{definition}[Empirical model] \label{def:empirical_model}%
	An empirical model $e$ on a measurement scenario $\XMO$ is a family of probability distributions $e = \left( e_C \right)_{C \in \mathcal{M}} $ where $e_C\colon O^{C} \to [0,1]$.
\end{definition}

Examples of empirical models are given in Table~\ref{tab:examples_empirical_model}.

\begin{table}[htpb]
	\centering
	\begin{subtable}[c]{0.5\textwidth}
		\centering
		\begin{tabular}[c]{@{} l l c c c c @{}}
			\toprule
			& & \multicolumn{4}{c}{$O^{C}$} \\ \cmidrule{3-6}
			\multicolumn{2}{c}{$C$} & $00$ & $01$ & $10$ & $11$ \\
			\midrule
			$A$ & $B$ & $\sfrac{1}{2}$ & $0$ & $0$ & $\sfrac{1}{2}$ \\
			$A$ & $B'$ & $\sfrac{1}{2}$ & $0$ & $0$ & $\sfrac{1}{2}$\\
			$A'$ & $B$ & $\sfrac{1}{2}$ & $0$ & $0$ & $\sfrac{1}{2}$\\
			$A'$ & $B'$ & $0$ &  $\sfrac{1}{2}$ & $\sfrac{1}{2}$ & $0$ \\
			\bottomrule
		\end{tabular}
		\caption{PR-Box~\cite{popescu1994QuantumNonlocalityAxiom}}
		\label{tab:example_empirical_PR}
	\end{subtable}%
	\begin{subtable}[c]{0.5\textwidth}
		\centering
		\begin{tabular}[c]{@{} l l c c c c @{}}
			\toprule
			& & \multicolumn{4}{c}{$O^{C}$} \\ \cmidrule{3-6}
			\multicolumn{2}{c}{$C$} & $00$ & $01$ & $10$ & $11$ \\
			\midrule
			$A$ & $B$ & $\sfrac{3}{5}$ & $0$ & $\sfrac{1}{5}$ & $\sfrac{1}{5}$ \\
			$A$ & $B'$ & $\sfrac{3}{5}$ & $0$ &  $\sfrac{1}{5}$ & $\sfrac{1}{5}$ \\
			$A'$ & $B$ & $\sfrac{3}{10}$ & $\sfrac{1}{5}$ & $\sfrac{1}{2}$ & $0$\\
			$A'$ & $B'$ & $\sfrac{3}{10}$ & $\sfrac{1}{5}$ & $\sfrac{1}{2}$ & $0$\\
			\bottomrule
		\end{tabular}
		\caption{Non-contextual empirical model}
		\label{tab:non-contextual_empirical_model_example}
	\end{subtable}
	\caption{Two examples of empirical models in the CHSH measurement scenario.}
	\label{tab:examples_empirical_model}
\end{table}

One of the main features we expect from empirical models is their \emph{compatibility}, which is a necessary requirement for non-contextuality:
\begin{definition}[Compatibility] \label{def:compatibility}%
An empirical model $e$ on a measurement scenario $\XMO$ is compatible if for every pair of contexts $C, C' \in \mathcal{M}$, the marginals agree on their overlap:
\begin{equation} \label{eq:compatibility_of_marginals}
	\restr{e_C}{C \cap C'} = \restr{e_{C'}}{C \cap C'}
.\end{equation}
Where the notation $\restr{e_C}{U}$ for all $U \subseteq C$ refers to the marginalization of probability distributions, i.e. for all $t \in O^{U}$
\begin{equation} \label{eq:marginal}
	\restr{e_C}{U}(t) \coloneq \sum_{s \in O^{C}, \restr{s}{U} = t} e_C(s) 
.\end{equation} 
\end{definition}

To summarize, the general setup of an experiment is given by a measurement scenario $\XMO$, then empirical data is gathered in a family of probability distributions given by $e = \left( e_C \right)_{C \in \mathcal{M}}$. 
 We now turn to the notion of non-contextuality, which is defined relative to a given empirical model $e$:
\begin{definition}[Non-contextuality] \label{def:non-contextuality}%
	An empirical model $e$ on a measurement scenario $\XMO$ is non-contextual if there exists a global probability distribution on the outcomes $d\colon O^{X} \to [0,1]$ such that the empirical model is obtained as its marginal:
	\begin{equation} \label{eq:non-contextuality-def}
		\forall C \in \mathcal{M}\colon e_C = \restr{d}{C}
	.\end{equation}
	If there is no such global distribution $d$, we say that the empirical model is contextual.
\end{definition}

For non-local scenarios, this notion of non-contextuality is equivalent to locality. It is also related to Kochen-Specker non-contextuality, through the existence of a factorizable hidden variable model.

\subsection{Contextual fraction} \label{ssec:contextual_fraction}

A measure for contextuality was derived in Ref~\cite{abramsky2017ContextualFractionMeasure}, called the \emph{Contextual Fraction} (CF). Given an empirical model, it is a bounded quantity, i.e. $0 \le \CF(e) \le 1$, that quantifies the deviation from a non-contextual empirical model.

More formally, we can write any empirical model as a convex combination of a non-contextual empirical model $e^{\text{NC}}$ and another empirical model $e'$ such that:
 \begin{equation} \label{eq:decomposition_empirical_model}
e = \lambda e^{\text{NC}} + (1 - \lambda) e'
.\end{equation} 
Where $\lambda \in [0,1]$. 
Let $\lambda^{*}$ be the maximum weight that can be assigned to the non-contextual empirical model $e^{\text{NC}}$, then we define the \emph{non-contextual fraction} as $\NCF(e) \coloneq \lambda^{*}$ and the contextual fraction as $\CF(e) \coloneq 1 - \NCF(e)$.

The contextual fraction can be computed with a linear program, implemented in \contextuality{}, which computes the largest non-contextual contribution to an empirical model $\mathbf{v}^{e}$:
\begin{equation} \tag{P-NCF} \label{lp:NCF}
	\begin{alignedat}{3}
		\max_{\mathbf{b}} &\quad \mathbf{1} \cdot \mathbf{b} \\
		\st &\quad \mathrm{M} \mathbf{b} \le \mathbf{v}^{e}\\
		    &\quad \mathbf{b} \ge \mathbf{0}
	.\end{alignedat}
\end{equation}
Where $\mathbf{b} \in \mathbb{R}^{n}$, $\mathrm{M}$ is the  \emph{incidence matrix}, $\mathbf{v}^{e}$ is the vectorized version of the empirical model $e$ and  $\mathbf{0}$ is the 0-vector. By definition, the NCF is given by the optimal value of the objective function: $\NCF(e) \coloneq \mathbf{1} \cdot \mathbf{b}^{*}$.
Information about the incidence matrix and other terms can be found in the original papers~\cite{abramsky2017ContextualFractionMeasure,abramsky2011SheaftheoreticStructureNonlocality}.

\subsection{Signalling fraction} \label{ssec:signalling_fraction}

Another measure, called the \emph{Signalling Fraction} (SF), was derived in~\cite{vallee2024CorrectedBellNoncontextuality}. It is also a bounded quantity, i.e. $0 \le \SF(e) \le 1$. Similarly to the CF, we can rewrite any empirical model $e$ as a convex combination of a compatible empirical model  $e^{\text{comp}}$ and any other empirical model $e''$:
 \begin{equation} \label{eq:NSF}
e = \gamma e^{\text{comp}} + \left( 1 - \gamma \right) e''
.\end{equation} 
With $\gamma \in [0,1]$. Once again, we define the \emph{non-signalling fraction} as the maximum value for $\gamma$ which we denote $\gamma^{*}$, i.e. $\NSF(e) \coloneq \gamma^{*}$ and the signalling fraction as $\SF(e) \coloneq 1 - \NSF(e)$. The signalling fraction can be formulated in terms of the following linear program:

\begin{equation} \tag{P-NSF} \label{lp:NSF}
	\begin{alignedat}{3}
		\max_{\mathbf{b}} &\quad \mathbf{1} \cdot \mathbf{b} \\
		\st &\quad \mathrm{M} \mathbf{b} \le \mathbf{v}^{e}\\
		    &\quad \mathrm{M} \mathbf{b} \ge \mathbf{0}
	.\end{alignedat}
\end{equation}
Where $\mathbf{b} \in \mathbb{R}^{n}$, $\mathrm{M}$ is the  \emph{incidence matrix}, $\mathbf{v}^{e}$ is the vectorized version of the empirical model $e$ and  $\mathbf{0}$ is the 0-vector. By definition, the NSF is given by the optimal value of the objective function: $\NSF(e) \coloneq \mathbf{1} \cdot \mathbf{b}^{*}$. The major difference with the contextual fraction linear program~\ref{lp:NCF} is the second constraint, which allows the vector $\mathbf{b}$ to take negative values.

\end{document}